\documentclass[12pt]{article}
\usepackage[margin=1in]{geometry}
\usepackage{newtxtext,newtxmath}
\usepackage{booktabs}
\usepackage{longtable}
\usepackage{array}
\usepackage{authblk}
\usepackage{hyperref}
\hypersetup{hidelinks}
\usepackage[super,numbers,sort&compress]{natbib} % superscript numeric citations
\usepackage{graphicx}


\title{Design-Based Validation Method of EHR Algorithms to Address Partial Verification Bias}
\author[1, 6]{Rushi Tang}
\author[2]{Maya Blasingame}
\author[2]{Lillian Zheng}
\author[2]{Joan Ac-Lumor}
\author[2]{Saher Mubarek}
\author[3, 4]{Jennifer K. Plichta}
\author[1, 5]{Samuel I. Berchuck}
\affil[1]{Department of Biostatistics and Bioinformatics, Duke University, Durham, NC}
\affil[2]{School of Medicine, Duke University, Durham, NC}
\affil[3]{Department of Surgery, Duke University, Durham, NC}
\affil[4]{Department of Population Health Sciences, Duke University, Durham, NC}
\affil[5]{Department of Statistical Science, Duke University, Durham, NC}
\affil[6]{Department of Population and Quantitative Health Sciences, Case Western Reserve University, Cleveland, OH}
\date{}

\begin{document}
\maketitle

\begin{abstract}
Electronic health record (EHR) phenotyping algorithms are increasingly used for clinical prediction, observational research, and quality improvement, but gold-standard validation by manual chart review is expensive. Many EHR algorithms target relatively rare clinical outcomes, so the number of algorithm-positive patients is often small while that of algorithm-negative patients could be orders of magnitude larger. We propose a design-based framework for validating such algorithms that fully verifies algorithm-positive patients and obtains a simple random sample (SRS) of algorithm-negative patients. Horvitz--Thompson (HT) estimators of false-negative (FN) and true-negative (TN) totals yield design-based estimates of sensitivity, specificity, negative predictive value (NPV), positive predictive value (PPV), and accuracy. We derive finite-population-corrected variance estimators and confidence intervals (CIs), and provide sample-size planning to target a prespecified CI half-width. An optional Phase~2 extension uses pilot data to determine whether risk-guided Neyman allocation may improve precision. In simulations, the design-based approach substantially reduced bias relative to naive verified-sample analyses and achieved near-nominal CI coverage. Adaptive allocation reduced root mean squared error (RMSE) and CI width when FNs were concentrated across risk strata, while offering little benefit when the risk signal was weak. We illustrate the framework using a breast cancer recurrence algorithm in 3,336 patients, with endpoint-specific algorithm-negative review budgets of 125, 150, and 175 charts. The framework combines SRS-based validation and precision planning with optional pilot-guided adaptive allocation.
\end{abstract}

\vspace{1em}

\section{Introduction}

Electronic health record (EHR) data are widely used in clinical research, safety surveillance, and quality improvement.\citep{Pendergrass2019,Richesson2016} Computable phenotypes use structured EHR data, such as diagnosis codes, procedures, laboratory test results, and medication records, to identify patients with specific diseases or clinical outcomes at scale.\citep{Richesson2016,Zigui2025,Kirby2016} Because these algorithms often define study cohorts or outcomes for downstream analyses, errors can introduce bias into clinical and epidemiological studies.\citep{Young2018,BeesleyMukherjee2022} Gold-standard validation is therefore needed before widespread use.\citep{Newton2013,Pendergrass2019} Clinician manual chart review is a common gold standard and can be used to estimate positive predictive value (PPV), negative predictive value (NPV), sensitivity, specificity, and accuracy. However, chart review is labor-intensive, and complete review is often infeasible in large EHR cohorts.\citep{Newton2013,Kirby2016}

Many EHR phenotype algorithms target relatively uncommon clinical outcomes. In these settings, the algorithm-positive group may be small enough for complete chart review, whereas the algorithm-negative group can be much larger. Investigators therefore often review all algorithm-positive patients to estimate PPV but review only a subset of algorithm-negative patients, or sometimes none at all.\citep{Rubbo2015,Swerdel2019} Because false negatives (FNs) occur in the algorithm-negative group, limited review makes sensitivity, specificity, NPV, and accuracy difficult to estimate precisely. In practice, these metrics are often omitted or estimated from small convenience samples without accounting for the sampling process or quantifying its uncertainty.\citep{CroninVickers2008,Bhasin2024} Given a fixed chart-review budget, investigators must decide which and how many algorithm-negative patients to review to obtain sufficiently precise operating-characteristic estimates. Practical guidance for making this design choice remains limited.

This setting involves partial verification: algorithm output is available for the full cohort, but the gold-standard outcome is observed only for patients selected for chart review.\citep{BeggGreenes1983,Zhou1998,Baker1995} Let \(T\) denote the algorithm output, \(D\) the true disease status, and \(V\) an indicator of gold-standard chart review. In EHR phenotype validation, \(T\) is observed for every patient, whereas \(D\) is observed only when \(V=1\). Partial verification bias arises when the reviewed patients do not represent the full cohort and the reviewed subset is analyzed as though it were complete. For example, if all algorithm-positive patients are reviewed but only a small, nonrandom subset of algorithm-negative patients is reviewed, the verified sample overrepresents algorithm-positive patients and may underrepresent FNs. Analyses based only on reviewed patients can therefore distort NPV, sensitivity, specificity, and accuracy. Most approaches to partial verification correct verification bias within missing-data frameworks, including methods for missing-at-random (MAR) and missing-not-at-random (MNAR) verification mechanisms.\citep{BeggGreenes1983,Zhou1993,ArifinYusof2022,KosinskiBarnhart2003,Lu2010} Many EHR validation studies instead sample patients from a finite cohort using a known design. In that setting, design-based methods that account directly for sampling probabilities may be more suitable for estimation and precision planning.

We propose a two-phase, design-based validation framework for EHR phenotyping studies. The framework applies when all algorithm-positive patients and a subset of algorithm-negative patients selected under a prespecified design undergo chart review. We derive Horvitz--Thompson (HT) and ratio estimators for finite-cohort operating characteristics and corresponding variance estimators with the finite-population correction. We also develop precision-based sample size planning for algorithm-negative review and an optional adaptive risk-guided sampling strategy. The adaptive strategy may improve review efficiency when FNs are concentrated across risk strata and Phase~1 data estimate this heterogeneity with sufficient stability. We evaluate the framework in simulations and apply it to a breast cancer recurrence detection algorithm using EHR data.

Section 2 reviews related work. Section 3 develops the validation framework. Section 4 evaluates finite-sample bias, RMSE, CI coverage, and adaptive-trigger behavior across disease prevalences, FN heterogeneity, algorithm-performance profiles, chart-review budgets, and pilot sizes; it also includes a sensitivity analysis with outcome-dependent Phase~2 verification. Section 5 applies the method to the breast cancer recurrence detection algorithm, and Section 6 concludes.

\section{Related Work}
Existing approaches to partial verification have largely focused on correcting verification bias within missing-data frameworks, assuming that verification is MAR: conditional on observed patient or test characteristics, the probability of gold-standard verification does not depend on unobserved disease status. \cite{BeggGreenes1983,Zhou1993,KosinskiBarnhart2003,Lu2010,ArifinYusof2022} Examples include Bayes-theorem adjustments and model-based mean-score, inverse probability weighting (IPW), and semiparametric efficient (SPE) estimators.\cite{BeggGreenes1983,Zhou1993,ArifinYusof2022} For example, IPW weights verified participants by the inverse of their estimated verification probability, yielding sensitivity and specificity estimators that are consistent under MAR and correct specification of the verification model. \cite{ArifinYusof2022} These methods extend to area under the receiver operating characteristic (AUROC), and corresponding R packages are available. Under MNAR, where verification remains associated with unobserved disease status after conditioning on observed information, more complex approaches include selection models that specify verification probability as a function of unobserved $D$ and likelihood-based estimators with sensitivity analyses.\cite{KosinskiBarnhart2003} Lu et al. also introduced a Bayesian method that accounts for both verification bias and imperfections in the gold standard.\cite{Lu2010} These approaches differ in their assumptions about verification and outcome models. Our focus is estimation and precision planning for a finite cohort sampled under a known design.

More recently, research has addressed both the design and analysis of efficient validation studies. In addition to obtaining valid estimates, investigators must decide how many charts to review and which charts to select to achieve adequate precision within a fixed review budget. Several studies have derived sample-size formulae and planning guidance under partial verification, often for sensitivity, specificity, or comparisons of two diagnostic tests. \cite{RoldanNofuentes2009,Shan2018} Adaptive and sequential validation designs have also been proposed, including multi-wave chart review with stopping rules and comparisons between sampling schemes such as simple random sampling (SRS) and stratified allocation. \cite{HahnAdaptive} Other approaches use auxiliary information to improve efficiency, including surrogate-assisted subsampling for valid logistic-regression coefficient inference and unsupervised coverage sampling to enhance chart-review coverage in EHR phenotype development.\cite{MarksAnglin2025, Zigui2025} Zou et al. used chart-reviewed data to estimate true-outcome probabilities conditional on recorded EHR outcomes and patient characteristics, then incorporated these probabilities into prediction-model validation. They also highlighted the need for efficient stratified and two-phase chart-review designs, motivating further work on how limited review budgets should be allocated.\citep{Zou2026} Together, these studies address complementary aspects of efficient validation, including sample selection and performance estimation.

Nevertheless, important gaps remain in the design of EHR phenotype validation studies. First, many studies are conducted in finite, well-defined EHR cohorts, with patients selected for chart review using probability sampling with known inclusion probabilities.\cite{HahnAdaptive,MarksAnglin2025} Established partial-verification methods, however, have largely been developed within superpopulations or missing-data frameworks.\cite{BeggGreenes1983,AlonzoPepe2005,ArifinYusof2022} Second, although reviewing all algorithm-positive patients and only a subset of algorithm-negative patients is common practice, investigators have little practical guidance for selecting algorithm-negative patients to attain a desired precision under a fixed review budget. Existing sampling approaches either focus on logistic-regression coefficient estimation or optimize one or two metrics, rather than offering a unified framework for estimating and planning precision across the full set of operating characteristics.\cite{HahnAdaptive,RoldanNofuentes2009,Shan2018,MarksAnglin2025} We address these gaps with a finite-population, two-phase validation framework that combines design-based estimation, precision-based sample-size planning, and adaptive risk-guided sampling for EHR algorithm validation.

\section{Methodology}
% :contentReference[oaicite:0]{index=0}
\subsection{Finite-population validation framework}

Let $\mathcal{U}=\{1,\ldots,N\}$ denote the finite EHR cohort after application of a phenotyping algorithm. For each patient $i \in \mathcal{U}$, let $P_i\in[0,1]$ denote the algorithm's predicted probability of the target outcome. Let $T_i \in \{0,1\}$ denote the corresponding binary algorithm classification, obtained by applying a prespecified classification rule to $P_i$. Here, $T_i=1$ indicates an algorithm-positive classification and $T_i=0$ an algorithm-negative classification. The cohort is partitioned into the algorithm-positive group \(\mathcal{U}_{+}=\{i:T_i=1\}\), with \(N_{+}=|\mathcal{U}_{+}|\), and the algorithm-negative group \(\mathcal{U}_{-}=\{i:T_i=0\}\), with \(N_{-}=|\mathcal{U}_{-}|\), where $N=N_{+}+N_{-}$. The gold standard is the chart-reviewed outcome status, denoted by $D_i \in \{0,1\}$, where $D_i=1$ indicates that the patient experienced the outcome and $D_i=0$ otherwise.

The goal of validation is to estimate the operating characteristics that would be obtained if the gold-standard outcome $D_i$ were known for every patient in the finite cohort. The corresponding finite-population cell counts are
\[
\mathrm{TP}=\sum_{i=1}^{N} I(T_i=1,D_i=1), \qquad
\mathrm{FP}=\sum_{i=1}^{N} I(T_i=1,D_i=0),
\]
\[
\mathrm{FN}=\sum_{i=1}^{N} I(T_i=0,D_i=1), \qquad
\mathrm{TN}=\sum_{i=1}^{N} I(T_i=0,D_i=0),
\]
where TP, FP, FN, and TN denote the numbers of true positives, false positives, false negatives, and true negatives, respectively. TP and FP occur among algorithm-positive patients, whereas FN and TN occur among algorithm-negative patients. Together, these four quantities define the algorithm's performance in the finite cohort. The operating characteristics of interest are
\[
\mathrm{PPV}=\frac{\mathrm{TP}}{\mathrm{TP}+\mathrm{FP}}
           =\frac{\mathrm{TP}}{N_{+}},
\qquad
\mathrm{NPV}=\frac{\mathrm{TN}}{\mathrm{TN}+\mathrm{FN}}
           =\frac{\mathrm{TN}}{N_{-}},
\]
\[
\mathrm{Sensitivity}=\frac{\mathrm{TP}}{\mathrm{TP}+\mathrm{FN}},
\qquad
\mathrm{Specificity}=\frac{\mathrm{TN}}{\mathrm{TN}+\mathrm{FP}},
\]
and
\[
\mathrm{Accuracy}=\frac{\mathrm{TP}+\mathrm{TN}}{N}.
\]
Once the four finite-population cell counts are known or estimated, all performance metrics follow directly.

In many EHR validation studies, the gold-standard outcome is not observed for every patient because manual chart review is costly. Let $V_i \in \{0,1\}$ indicate whether patient $i$ is selected for chart review. When $V_i=1$, the chart-reviewed outcome $D_i$ is observed; when $V_i=0$, it remains unobserved. The algorithm classification $T_i$ and the cohort sizes $N_{+}$ and $N_{-}$ are known for all patients before chart review, whereas outcome status is unknown for unreviewed patients.

The target cohort is fixed and fully enumerated before validation, so $\mathrm{TP}$, $\mathrm{FP}$, $\mathrm{FN}$, and $\mathrm{TN}$ are fixed but partially unobserved totals. Chart review is treated as a probability sample from this cohort, and inference is based on the known design used to select patients for verification. We assume that chart review completely and correctly ascertains the gold-standard outcome \(D_i\) for every reviewed patient, with no missing or misclassified outcome information.

We begin with the simplest validation design: all algorithm-positive patients are reviewed, and a simple random sample without replacement (SRSWOR) of algorithm-negative patients is selected for chart review. Hereafter, SRS denotes SRSWOR unless otherwise stated.

%\paragraph{Design assumptions.}
%The framework targets a fixed, fully enumerated EHR cohort in which the algorithm predictions \(P_i\), binary classifications \(T_i\), and cohort membership are known before chart review. We assume that chart review provides the prespecified gold-standard outcome \(D_i\) for reviewed patients. Validity of the design-based estimators requires that the chart-review design be implemented as specified and that each patient eligible for review have a known, positive inclusion probability under the realized design. Under the adaptive design, the Phase~2 allocation may depend on the realized Phase~1 sample and outcomes, but, conditional on this information, Phase~2 selection follows the known sampling probabilities within the prespecified strata.

%These assumptions do not require a missing-at-random model for unreviewed outcomes or correct calibration of the predicted risks. The risk score is used only to define strata, plan the allocation, and potentially improve precision; its informativeness affects the potential efficiency gain from adaptive sampling, not design-based validity. The delta-method and logit-scale CIs additionally rely on sufficiently stable variance estimation. The exact finite-population boundary procedure in Section~3.5 applies when no FNs are observed under the final SRS design.

\subsection{Estimation under SRS}

We first consider the simplest validation design. All algorithm-positive patients are reviewed, and an SRS of $m$ algorithm-negative patients is selected from $\mathcal{U}_{-}$. Let $S_{-} \subset \mathcal{U}_{-}$ denote the sampled set, with $|S_{-}|=m$. Under SRS, each algorithm-negative patient has sampling probability
$
f=m/N_{-}
$,
where \(f\) is the sampling fraction among algorithm-negative patients. The common sampling weight is
$w=1/f=N_{-}/m$.

Because all algorithm-positive patients are verified, their cell counts are observed directly:
\(a=\sum_{i \in \mathcal{U}_{+}} I(D_i=1)\) and
\(b=\sum_{i \in \mathcal{U}_{+}} I(D_i=0)\),
where $a$ is the number of TP and $b$ is the number of FP among algorithm-positive patients. Thus, $N_{+}=a+b$, and no sampling adjustment is needed for this group.

The FN and TN totals among algorithm-negative patients are unknown. Because only a sample of this group is reviewed, we estimate these totals with HT estimators. Let
\(c=\sum_{i \in S_{-}} I(D_i=1)\) and
\(d=\sum_{i \in S_{-}} I(D_i=0)\)
denote the observed numbers of FNs and TNs in the reviewed negative sample, so $m=c+d$. The HT estimators of the finite-population totals are
$
\widehat{\mathrm{FN}}_{\mathrm{SRS}}
=
\sum_{i \in S_{-}} I(D_i=1)/f
=
wc$,
and
$
\widehat{\mathrm{TN}}_{\mathrm{SRS}}
=
\sum_{i \in S_{-}} I(D_i=0)/f
=
wd$.
These estimators inflate the observed sample counts by the inverse of the sampling fraction. Under SRS, they are design-unbiased for the corresponding finite-population $\mathrm{FN}$ and $\mathrm{TN}$ totals.

We estimate the operating characteristics by substituting the observed positive-side counts and HT estimates of the negative-side totals into the complete-cohort definitions. PPV is observed directly because all algorithm-positive patients are reviewed:
\(\mathrm{PPV}=a/(a+b)=a/N_{+}\).
The remaining operating characteristics are estimated as
\begin{equation}
\begin{aligned}
\widehat{\mathrm{NPV}}_{\mathrm{SRS}}
&=
\frac{\widehat{\mathrm{TN}}_{\mathrm{SRS}}}{N_-}
=
\frac{d}{m},
&
\widehat{\mathrm{Sens}}_{\mathrm{SRS}}
&=
\frac{a}{a+\widehat{\mathrm{FN}}_{\mathrm{SRS}}}, \\
\widehat{\mathrm{Spec}}_{\mathrm{SRS}}
&=
\frac{\widehat{\mathrm{TN}}_{\mathrm{SRS}}}{\widehat{\mathrm{TN}}_{\mathrm{SRS}}+b},
&
\widehat{\mathrm{Acc}}_{\mathrm{SRS}}
&=
\frac{a+\widehat{\mathrm{TN}}_{\mathrm{SRS}}}{N}.
\end{aligned}
\label{eq:operating_characteristic_estimators}
\end{equation}

We use the same substitution approach under the adaptive design, estimating FN and TN totals before calculating the operating characteristics.

\subsection{Variance estimation and confidence intervals (CIs)}

We next derive design-based variance estimators for the operating-characteristic
estimators under SRS. The sampling fraction is \(f=m/N_{-}\). Define the
finite-population proportions \(p_{\mathrm{FN}}=\mathrm{FN}/N_{-}\) and
\(p_{\mathrm{TN}}=\mathrm{TN}/N_{-}=1-p_{\mathrm{FN}}\). Under SRS,
their sample estimates are \(\widehat p_{\mathrm{FN}}=c/m\) and
\(\widehat p_{\mathrm{TN}}=d/m\).

Under SRS, the observed number of FNs in the reviewed algorithm-negative sample follows a hypergeometric distribution. Conditional on the fixed finite-population FN total,
$
c \sim \operatorname{Hypergeometric}(N_{-},\mathrm{FN},m).
$
Therefore,
\[
\operatorname{Var}(c)
=
m p_{\mathrm{FN}}(1-p_{\mathrm{FN}})
\frac{N_{-}-m}{N_{-}-1}.
\]
Dividing by \(m^2\) gives
\[
\operatorname{Var}(\widehat p_{\mathrm{FN}})
=
\left(1-f\right)
\frac{N_{-}}{N_{-}-1}
\frac{p_{\mathrm{FN}}(1-p_{\mathrm{FN}})}{m}
=
\left(1-f\right)\frac{S_{\mathrm{FN}}^2}{m},
\]
where \(S_{\mathrm{FN}}^2=\{N_{-}/(N_{-}-1)\}p_{\mathrm{FN}}(1-p_{\mathrm{FN}})\) is the finite-population variance of the binary FN indicator. The finite-population correction \(1-f\) arises because sampling is without replacement. As the reviewed sample approaches the full algorithm-negative cohort, sampling uncertainty vanishes.

Replacing the unknown finite-population variances with their SRS sample estimators gives
\[
\widehat{\mathrm{Var}}(\widehat p_{\mathrm{FN}})
=
(1-f)\frac{\widehat p_{\mathrm{FN}}(1-\widehat p_{\mathrm{FN}})}{m-1},
\qquad
\widehat{\mathrm{Var}}(\widehat p_{\mathrm{TN}})
=
(1-f)\frac{\widehat p_{\mathrm{TN}}(1-\widehat p_{\mathrm{TN}})}{m-1}.
\]
The same result applies to \(\widehat p_{\mathrm{TN}}=d/m\), since the number of TNs is \(d=m-c\).

Because \(\widehat{\mathrm{FN}}_{\mathrm{SRS}}=N_{-}\widehat p_{\mathrm{FN}}\)
and \(\widehat{\mathrm{TN}}_{\mathrm{SRS}}=N_{-}\widehat p_{\mathrm{TN}}\),
the corresponding design-based variance estimators for the finite-population
totals are
$
\widehat{\mathrm{Var}}(\widehat{\mathrm{FN}}_{\mathrm{SRS}})
=
N_{-}^{2}\widehat{\mathrm{Var}}(\widehat p_{\mathrm{FN}}),
$
and
$
\widehat{\mathrm{Var}}(\widehat{\mathrm{TN}}_{\mathrm{SRS}})
=
N_{-}^{2}\widehat{\mathrm{Var}}(\widehat p_{\mathrm{TN}}).
$
Treating the fully observed positive-side counts \(a\) and \(b\) as fixed gives the following delta-method variance estimators:
\begin{equation}
\begin{aligned}
\widehat{\mathrm{Var}}(\widehat{\mathrm{NPV}}_{\mathrm{SRS}})
&=
\frac{\widehat{\mathrm{Var}}(\widehat{\mathrm{TN}}_{\mathrm{SRS}})}{N_-^2},
&
\widehat{\mathrm{Var}}(\widehat{\mathrm{Sens}}_{\mathrm{SRS}})
&=
\left\{\frac{a}{(a+\widehat{\mathrm{FN}}_{\mathrm{SRS}})^2}\right\}^{2}
\!\widehat{\mathrm{Var}}(\widehat{\mathrm{FN}}_{\mathrm{SRS}}), \\
\widehat{\mathrm{Var}}(\widehat{\mathrm{Spec}}_{\mathrm{SRS}})
&=
\left\{\frac{b}{(\widehat{\mathrm{TN}}_{\mathrm{SRS}}+b)^2}\right\}^{2}
\!\widehat{\mathrm{Var}}(\widehat{\mathrm{TN}}_{\mathrm{SRS}}),
&
\widehat{\mathrm{Var}}(\widehat{\mathrm{Acc}}_{\mathrm{SRS}})
&=
\frac{\widehat{\mathrm{Var}}(\widehat{\mathrm{TN}}_{\mathrm{SRS}})}{N^2}.
\end{aligned}
\label{eq:delta_method_variances}
\end{equation}

PPV is exact, with no sampling uncertainty, because all algorithm-positive patients are reviewed.
If the validated algorithm-positive group is viewed as representing a broader
population of possible algorithm-positive patients, uncertainty in PPV can be
summarized using the usual binomial variance,
\(\widehat{\mathrm{Var}}(\widehat{\mathrm{PPV}})=
\widehat{\mathrm{PPV}}(1-\widehat{\mathrm{PPV}})/(a+b)\).
In the analyses below, we report Wilson score CIs for PPV based on $a$ TPs among $a+b$ reviewed algorithm-positive patients.

For NPV, sensitivity, specificity, and accuracy, we use a logit-transformed Wald procedure with the corresponding design-based variance estimator. For a generic operating characteristic $\theta \in (0,1)$ with estimator $\widehat{\theta}$ and variance estimator $\widehat{\mathrm{Var}}(\widehat{\theta})$, the approximate logit-scale standard error is
\begin{equation}
\widehat{\mathrm{SE}}\{\mathrm{logit}(\widehat{\theta})\}
=
\frac{\sqrt{\widehat{\mathrm{Var}}(\widehat{\theta})}}
     {\widehat{\theta}(1-\widehat{\theta})}.
\label{eq:logit_se}
\end{equation}
A two-sided $(1-\alpha)$ CI is then computed as
\begin{equation}
\mathrm{logit}(\widehat{\theta})
\pm
z_{1-\alpha/2}
\widehat{\mathrm{SE}}\{\mathrm{logit}(\widehat{\theta})\}.
\label{eq:logit_ci}
\end{equation}
The interval limits are then transformed back to the probability scale using the inverse-logit transformation. This approach keeps the endpoints within $[0,1]$ while incorporating the finite-population sampling variance induced by reviewing only a subset of algorithm-negative patients.

\subsection{Sample size planning}

Before chart review, investigators must determine how many algorithm-negative patients to review. Sample-size planning concerns the total number of algorithm-negative patients, denoted by \(m\), sampled from \(\mathcal U_-\). Sample-size planning quantifies the trade-off between chart-review burden and anticipated precision.

We use SRS as the reference design for selecting \(m\). This conservative planning basis does not assume that predicted risk scores identify meaningful heterogeneity in FN rates or that the adaptive sampling rule in Section~3.5 will be triggered. For a candidate sample size \(m\), the sampling fraction among algorithm-negative patients is $f(m)=m/N_-.$

Let \(\widetilde p_{\mathrm{FN}}\) denote the planning FN
proportion in \(\mathcal U_-\), and let
\(\widetilde p_{\mathrm{TN}}=1-\widetilde p_{\mathrm{FN}}\).
Throughout this subsection, a tilde indicates a planning value or projected quantity. For example, $\widetilde{\operatorname{Var}}\{\cdot\}$
denotes a projected design variance for sample-size planning.

For a binary outcome under SRS, the projected variances follow directly from the SRS variance derived in Section~3.3. Before chart review, the finite-population FN proportion is unknown. We therefore replace \(p_{\mathrm{FN}}\) and \(p_{\mathrm{TN}}\) with the planning values \(\widetilde p_{\mathrm{FN}}\) and \(\widetilde p_{\mathrm{TN}}=1-\widetilde p_{\mathrm{FN}}\). Thus,
\[
\widetilde{\operatorname{Var}}
\left\{
\widehat p_{\mathrm{FN}}(m)
\right\}
=
\left(1-\frac{m}{N_-}\right)
\frac{N_-}{N_- - 1}
\frac{
\widetilde p_{\mathrm{FN}}
\left(1-\widetilde p_{\mathrm{FN}}\right)
}{m},
\]
and
\[
\widetilde{\operatorname{Var}}
\left\{
\widehat p_{\mathrm{TN}}(m)
\right\}
=
\left(1-\frac{m}{N_-}\right)
\frac{N_-}{N_- - 1}
\frac{
\widetilde p_{\mathrm{TN}}
\left(1-\widetilde p_{\mathrm{TN}}\right)
}{m}.
\]
The first factor is the finite-population correction described in Section~3.3. The factor \(N_-/(N_- - 1)\) expresses the finite-population variance of a binary outcome in terms of its proportion. Unlike the estimation formula in Section~3.3, which uses a sample-based variance estimator, the planning formula uses the projected finite-population proportion directly.
As \(\widehat{\mathrm{FN}}_{\mathrm{SRS}}(m)=
N_-\widehat p_{\mathrm{FN}}(m)\) and
\(\widehat{\mathrm{TN}}_{\mathrm{SRS}}(m)=
N_-\widehat p_{\mathrm{TN}}(m)\),
the projected variances of the estimated finite-population totals are
$
\widetilde{\operatorname{Var}}
\left\{
\widehat{\mathrm{FN}}_{\mathrm{SRS}}(m)
\right\}
=
N_-^2
\widetilde{\operatorname{Var}}
\left\{
\widehat p_{\mathrm{FN}}(m)
\right\},
$
and
$
\widetilde{\operatorname{Var}}
\left\{
\widehat{\mathrm{TN}}_{\mathrm{SRS}}(m)
\right\}
=
N_-^2
\widetilde{\operatorname{Var}}
\left\{
\widehat p_{\mathrm{TN}}(m)
\right\}.
$

Planning values can be obtained from the algorithm's predicted outcome probabilities, which are computed before chart review. Using $P_i$, the predicted probability of the target outcome for patient \(i\), we approximate the expected positive-side cell counts by
$
\widetilde a
=
\sum_{i\in \mathcal U_+} P_i$, 
$\widetilde b
=
N_+-\widetilde a$
and the expected negative-side cell counts by $\widetilde{\mathrm{FN}}
=
\sum_{i\in \mathcal U_-} P_i$,
$\widetilde{\mathrm{TN}}
=
N_- - \widetilde{\mathrm{FN}}$.

These quantities imply
\(\widetilde p_{\mathrm{FN}}=\widetilde{\mathrm{FN}}/N_-\) and
\(\widetilde p_{\mathrm{TN}}=\widetilde{\mathrm{TN}}/N_-\).
The predicted risks are used only to obtain planning values before chart review. For each candidate value of \(m\), planning values of NPV, sensitivity, specificity, and accuracy are obtained by substituting \(\widetilde a\), \(\widetilde b\), \(\widetilde{\mathrm{FN}}\), and \(\widetilde{\mathrm{TN}}\) into Equation~\eqref{eq:operating_characteristic_estimators}. Their projected variances are obtained from Equation~\eqref{eq:delta_method_variances} by replacing the finite-population totals and variance estimators with the planning counterparts defined above.

The precision of \(\widehat{\mathrm{PPV}}=a/(a+b)\) does not depend on the number of algorithm-negative patients reviewed, and thus PPV does not contribute to the selection of \(m\) under the present design.

We evaluate projected precision over a prespecified grid of feasible sample sizes,
$
\mathcal M
=
\left\{
m^{(1)},m^{(2)},\ldots,m^{(L)}
\right\}.
$
Candidate values should reflect realistic chart-review workloads. Superscripts distinguish the candidate total sample sizes from the Phase~1 and Phase~2 sample sizes \(m_1\) and \(m_2\) introduced in Section~3.5.

For each \(m\in\mathcal M\) and operating characteristic \(\theta\), let \(\widetilde\theta\) denote its planning value and let $\widetilde{\operatorname{Var}}
\left\{
\widehat\theta_{\mathrm{SRS}}(m)
\right\}$ denote its projected variance. To match the final analysis, we evaluate the logit-scale standard-error and CI constructions in Equations~\eqref{eq:logit_se} and \eqref{eq:logit_ci} after replacing \(\widehat\theta\) and \(\widehat{\operatorname{Var}}(\widehat\theta)\) with \(\widetilde\theta\) and $\widetilde{\operatorname{Var}}\{\widehat\theta_{\mathrm{SRS}}(m)\}$, respectively. Let \(\widetilde L_\theta(m)\) and \(\widetilde U_\theta(m)\) denote the resulting inverse-logit-transformed lower and upper limits.
The projected CI half-width is then
$
\widetilde h_\theta(m)
=
[\widetilde U_\theta(m)-\widetilde L_\theta(m)]/2.
$

Investigators may specify a set of priority operating characteristics \(\mathcal K\), such as sensitivity and NPV, together with target half-widths \(h_k^*\) for each \(k\in\mathcal K\). A practical total sample size is the smallest candidate satisfying all prespecified precision requirements:
\[
m^*
=
\min
\left\{
m\in\mathcal M:
\widetilde h_k(m)\leq h_k^*
\text{ for every }k\in\mathcal K
\right\}.
\]
If no candidate sample size satisfies all target half-widths, the planning table still shows the incremental precision gained by increasing chart-review effort. Investigators can then select a feasible sample size by balancing anticipated precision against the practical cost of manual chart review.

% Because projected precision depends on the planning values, sensitivity analyses should be conducted when the predicted risks are uncertain or potentially miscalibrated. For example, the calculations may be repeated over a plausible range of false-negative proportions. Investigators may then select \(m\) using the scenario most relevant to the primary scientific objective or a prespecified conservative scenario.

The resulting \(m\) defines the total algorithm-negative chart-review budget for either SRS or the adaptive two-phase design. After selecting \(m\), investigators may evaluate the adaptive design parameters, including the Phase~1 size \(m_1\), number of risk strata \(H\), and triggering threshold \(\rho\), through simulation under plausible risk--outcome relationships. This evaluation determines how to deploy the fixed chart-review budget. The adaptive procedure in Section~3.5 extends the planning framework while preserving SRS as the reference design for total sample-size planning.

\subsection{Adaptive two-phase validation design}

SRS provides a transparent baseline for finite-population validation. However, when FNs are concentrated among algorithm-negative patients with higher predicted risks, SRS may devote substantial review effort to patients who are unlikely to be FNs. We therefore extend the basic design to an adaptive two-phase design that uses an initial pilot sample to determine whether risk-guided allocation is expected to improve the precision of sensitivity estimation.

Using pilot data to estimate an efficient Phase~2 allocation is motivated by a broader literature on optimal and adaptive two-phase sampling. Earlier work showed how pilot information can estimate Phase~2 sampling fractions under resource constraints.\cite{Reilly1996} Subsequent adaptive two-phase designs used internal pilot samples to estimate the nuisance quantities needed for approximately optimal validation allocation.\cite{McIsaacCook2015,Han2021} Related multi-wave procedures have also been developed for EHR validation and chart-review studies.\cite{HahnAdaptive} These methods support pilot-guided adaptive allocation, but generally target regression parameters, stopping decisions, or other validation quantities. Here, we adapt this principle to finite-population estimation of operating characteristics, using the allocation to reduce sensitivity variance through more precise estimation of the FN total.

Let \(m\) denote the total number of algorithm-negative patients to be reviewed, as determined by the sample-size planning procedure in Section~3.4. The total sample is divided into a Phase~1 pilot sample of size \(m_1\) and a Phase~2 sample of size $m_2=m-m_1$.

In Phase~1, an SRS of \(m_1\) patients is drawn from \(\mathcal U_-\), and their charts are reviewed. Let \(\mathcal S_1\subset \mathcal U_-\) denote the Phase~1 sample. The quantity \(P_i\) is the predicted outcome probability available before chart review. When the binary algorithm classification \(T_i\) is obtained by thresholding a predicted probability, \(P_i\) is the corresponding continuous prediction before thresholding.

Using \(P_i\), the algorithm-negative cohort is partitioned into \(H\) prespecified risk strata. Let \(\mathcal U_{-h}\) denote stratum \(h\), with \(N_h=|\mathcal U_{-h}|\), so \(\sum_{h=1}^{H}N_h=N_-\).
For example, strata may be defined by quantiles of \(P_i\) among algorithm-negative patients. Their definitions are fixed before the Phase~1 gold-standard outcomes are examined.

Let \(\mathcal S_{1h}=\mathcal S_1\cap \mathcal U_{-h}\) denote the Phase~1
sample in stratum \(h\), and define \(m_{1h}=|\mathcal S_{1h}|\) and
\(c_{1h}=\sum_{i\in\mathcal S_{1h}}D_i\), where $D_i$ is known from Phase~1. Then \(c_{1h}\) is the observed
number of FNs in stratum \(h\), and
\(c_1=\sum_{h=1}^{H}c_{1h}\) is the total number of FNs observed
in Phase~1.

We define the Jeffreys-type stabilized pilot estimate in stratum \(h\) as
\[
\widetilde p_{\mathrm{FN},h}^{(1)}
=
\begin{cases}
\dfrac{c_{1h}+1/2}{m_{1h}+1},
& m_{1h}>0,\\[8pt]
\dfrac{c_1+1/2}{m_1+1},
& m_{1h}=0.
\end{cases}
\]
For \(m_{1h}>0\), this is the posterior mean under a \(\mathrm{Beta}(1/2,1/2)\) Jeffreys prior in a binomial working model.\citep{Jeffreys1946,BrownCaiDasGupta2001} When \(m_{1h}=0\), we use the stabilized overall pilot estimate as a prespecified pooling fallback because stratum-specific pilot information is unavailable.
The corresponding pilot-estimated within-stratum standard deviation (SD) of the FN indicator is
$
\widetilde S_h
=
\left[
\widetilde p_{\mathrm{FN},h}^{(1)}
\left\{
1-\widetilde p_{\mathrm{FN},h}^{(1)}
\right\}
\right]^{1/2}$.
The Jeffreys-type adjustment prevents strata with few observed FNs from having an estimated within-stratum variance of zero. The pooling fallback gives every stratum a defined Phase~2 allocation. These quantities are used only for the adaptive decision and Phase~2 allocation.

\subsubsection{Sensitivity-based adaptive decision rule}

We use sensitivity as the primary operating characteristic for deciding whether adaptive allocation is worthwhile. Because
\(\mathrm{Sens}=a/(a+\mathrm{FN})\),
and the positive-side count \(a\) is fully observed, sampling uncertainty in sensitivity arises only from estimation of the FN total. Reducing the design variance of the estimated FN total therefore also reduces the variance of the sensitivity estimator.

Define the pilot-based overall FN proportion as
\[
\widetilde p_{\mathrm{FN}}^{(1)}
=
\sum_{h=1}^{H}
\frac{N_h}{N_-}
\widetilde p_{\mathrm{FN},h}^{(1)}.
\]
The Phase~1 projected variance of the FN total under continued SRS to total sample size \(m\) is
\[
\widehat{\operatorname{Var}}^{(1)}
\left\{
\widehat{\mathrm{FN}}_{\mathrm{SRS}}(m)
\right\}
=
N_-^2
\left(1-\frac{m}{N_-}\right)
\frac{
\widetilde p_{\mathrm{FN}}^{(1)}
\left\{
1-\widetilde p_{\mathrm{FN}}^{(1)}
\right\}
}{
m-1
}.
\]
Here, $\widehat{\operatorname{Var}}^{(1)}\{\cdot\}$
denotes a variance projected from the Phase~1 pilot data, not a final variance estimate calculated after both sampling phases are complete.

Under the provisional adaptive design, the \(m_2\) remaining chart reviews are allocated using a plug-in Neyman allocation. This allocation distributes a fixed review budget across strata to minimize the projected variance of the estimated FN total:
\[
m_{2h}^{*}
=
\operatorname{round}
\left[
m_2
\frac{
N_h\widetilde S_h
}{
\sum_{k=1}^{H}N_k\widetilde S_k
}
\right].
\]
Integer rounding is adjusted to ensure that
\(\sum_{h=1}^{H}m_{2h}^{*}=m_2\).
The allocation is also constrained by the number of unreviewed patients remaining in each stratum. When the review budget permits, each nonempty remaining stratum receives at least two Phase~2 observations so that its conditional variance can be estimated. If these constraints cannot be met for the selected \(H\), the design defaults to continued SRS.

The projected final sample size in stratum \(h\) is
\(m_h^{*}=m_{1h}+m_{2h}^{*}\).
Using the stabilized pilot estimates, the projected variance of the FN total under the provisional Neyman allocation is
\[
\widehat{\operatorname{Var}}^{(1)}
\left\{
\widehat{\mathrm{FN}}_{\mathrm{Ney}}(m)
\right\}
=
\sum_{h=1}^{H}
N_h^2
\left(1-\frac{m_h^{*}}{N_h}\right)
\frac{
\widetilde p_{\mathrm{FN},h}^{(1)}
\left\{
1-\widetilde p_{\mathrm{FN},h}^{(1)}
\right\}
}{
m_h^{*}-1
}.
\]
This projected variance is used only to determine the Phase~2 sampling design. Final estimation and inference explicitly account for the sequential two-phase sampling process.

The adaptive decision is based on the projected relative variance of the sensitivity estimator. Because the same pilot-based delta-method derivative maps the estimated FN total to sensitivity under both candidate allocations, this ratio is equivalently
\[
\widehat R(m)
=
\frac{
\widehat{\operatorname{Var}}^{(1)}
\left\{
\widehat{\mathrm{FN}}_{\mathrm{SRS}}(m)
\right\}
}{
\widehat{\operatorname{Var}}^{(1)}
\left\{
\widehat{\mathrm{FN}}_{\mathrm{Ney}}(m)
\right\}
}.
\]
Thus, although \(\widehat R(m)\) is calculated from projected variances of the FN total, it represents the projected relative variance of the sensitivity estimator.

A value of \(\widehat R(m)>1\) indicates that the provisional Neyman allocation is expected to reduce sensitivity variance relative to continued SRS. To avoid switching to a more complex adaptive design for a negligible projected gain, we prespecify a threshold \(\rho>1\). This user-specified threshold represents the minimum projected efficiency gain considered meaningful enough to justify adaptive allocation. Larger values of \(\rho\) favor continued SRS and require stronger Phase~1 evidence before switching designs. Adaptive allocation is triggered when \(\widehat R(m)>\rho\). In that case, \(m_{2h}=m_{2h}^{*}\) patients are sampled by SRS from the unreviewed patients in stratum \(h\). Let
$
\mathcal S_{2h}
\subset
\mathcal U_{-h}\setminus\mathcal S_{1h}
$
denote the Phase~2 sample in stratum \(h\), and let
$
c_{2h}
=
\sum_{i\in\mathcal S_{2h}}D_i
$
denote the number of FNs observed in that sample. If adaptive allocation is not triggered, the remaining \(m_2\) patients are sampled by SRS from \(\mathcal U_-\setminus\mathcal S_1\). 

\subsubsection{Composite estimator}

The adaptive estimator is based on design-based estimation for two-phase and sequential probability samples. The HT principle gives unbiased estimation of a finite-population total when the relevant inclusion probabilities are known.\cite{HorvitzThompson1952} More generally, two-phase sampling can be analyzed by conditioning on the realized Phase~1 sample and applying Phase~2 sampling weights to units that remain subject to selection.\cite{SarndalSwensson1987,SalehiSeber2001} For final inference, we use a fixed-weight composite estimator constructed from a sequential conditional HT component and the Phase~1 SRS estimator. Composite estimation has a long history in survey sampling as a way to combine information from different samples or sampling occasions.\cite{RaoGraham1964}

After Phase~1, the outcomes of the \(m_{1h}\) pilot patients in stratum \(h\) are observed and do not require weighting in the Phase~2 analysis. Let
$
M_h=N_h-m_{1h}
$
denote the number of patients remaining in stratum \(h\) after Phase~1. Conditional on the realized Phase~1 sample, each remaining patient in stratum \(h\) has Phase~2 inclusion probability
$
m_{2h}/M_h.
$
The sequential conditional HT component estimator of the FN total is
\[
\widehat{\mathrm{FN}}_{\mathrm{seq}}
=
\sum_{h=1}^{H}
\left(
c_{1h}
+
\frac{M_h}{m_{2h}}c_{2h}
\right).
\]
The first term retains the observed Phase~1 FNs, and the second estimates FNs among patients unreviewed after Phase~1. Conditional on the realized Phase~1 sample and resulting Phase~2 allocation,
$
\mathbb{E}_2
\left(
\widehat{\mathrm{FN}}_{\mathrm{seq}}
\mid\mathcal S_1
\right)
=
\mathrm{FN},
$
where \(\mathbb{E}_2\) denotes expectation over the Phase~2 sampling design. Thus, \(\widehat{\mathrm{FN}}_{\mathrm{seq}}\) is conditionally design-unbiased and, by iterated expectation, design-unbiased over the joint sampling design. However, when FNs are rare, one or more Phase~2 strata may contain few or no observed FNs. The resulting stratum-specific plug-in variance estimates can be unstable or degenerate, producing unreliable CIs when the sequential component is used alone.

Let
$
\widehat{\mathrm{FN}}_1
=
\frac{N_-}{m_1}c_1
$
denote the Phase~1 SRS estimator, where \(c_1=\sum_{h=1}^{H}c_{1h}\). Our primary adaptive estimator is
$
\widehat{\mathrm{FN}}_{\mathrm C}
=
\lambda\widehat{\mathrm{FN}}_1
+
(1-\lambda)\widehat{\mathrm{FN}}_{\mathrm{seq}},
$
where
\[
\lambda
=
\frac{
m_1(N_--m)
}{
m(N_--m_1)
}.
\]
The fixed weight is derived for the present two-phase design and ensures coherence with the basic SRS estimator. If the adaptive rule is not triggered, Phase~2 is an SRS from all \(N_--m_1\) remaining algorithm-negative patients, and the composite estimator reduces exactly to
\[
\widehat{\mathrm{FN}}_{\mathrm C}
=
\frac{N_-}{m}(c_1+c_2),
\]
where \(c_2\) is the number of FNs observed in the Phase~2 SRS sample. This is the ordinary SRS HT estimator based on all \(m\) reviewed algorithm-negative patients.

The weight \(\lambda\) depends only on prespecified population and sample sizes, not on observed outcomes. Because both component estimators are design-unbiased and \(\lambda\) is fixed by design, \(\widehat{\mathrm{FN}}_{\mathrm C}\) is design-unbiased over the joint Phase~1 and Phase~2 sampling process. The corresponding TN total is
$
\widehat{\mathrm{TN}}_{\mathrm C}
=
N_--\widehat{\mathrm{FN}}_{\mathrm C}.
$

\paragraph{Variance estimation.}

Let $\widehat p_{\mathrm{FN}}^{(1)}
=c_1/m_1$.
The design-based variance estimator for the Phase~1 estimator is
\[
\widehat{\operatorname{Var}}
\left\{
\widehat{\mathrm{FN}}_1
\right\}
=
N_-^2
\left(1-\frac{m_1}{N_-}\right)
\frac{
\widehat p_{\mathrm{FN}}^{(1)}
\left\{
1-\widehat p_{\mathrm{FN}}^{(1)}
\right\}
}{
m_1-1
}.
\]

For the triggered adaptive design, define
$
\widehat p_{\mathrm{FN},2h}
=
c_{2h}/m_{2h}
$.
Conditional on the Phase~1 sample, the variance estimator for the sequential conditional estimator is
\[
\widehat{\operatorname{Var}}
\left\{
\widehat{\mathrm{FN}}_{\mathrm{seq}}
\mid\mathcal S_1
\right\}
=
\sum_{h=1}^{H}
M_h^2
\left(1-\frac{m_{2h}}{M_h}\right)
\frac{
\widehat p_{\mathrm{FN},2h}
\left\{
1-\widehat p_{\mathrm{FN},2h}
\right\}
}{
m_{2h}-1
}.
\]
A stratum that is completely reviewed in Phase~2 has \(m_{2h}=M_h\) and therefore contributes zero conditional variance.

When the adaptive rule is not triggered, the corresponding conditional variance estimator is
\[
\widehat{\operatorname{Var}}
\left\{
\widehat{\mathrm{FN}}_{\mathrm{seq}}
\mid\mathcal S_1
\right\}
=
(N_--m_1)^2
\left(
1-\frac{m_2}{N_--m_1}
\right)
\frac{
\widehat p_{\mathrm{FN}}^{(2)}
\left\{
1-\widehat p_{\mathrm{FN}}^{(2)}
\right\}
}{
m_2-1
},
\]
where
$\widehat p_{\mathrm{FN}}^{(2)}
=c_2/m_2$. The covariance between the two component estimators is zero under the joint sampling design. By the law of total covariance,
\[
\begin{aligned}
\operatorname{Cov}_{1,2}
\left(
\widehat{\mathrm{FN}}_1,
\widehat{\mathrm{FN}}_{\mathrm{seq}}
\right)
&=
\mathbb{E}_1
\left[
\operatorname{Cov}_2
\left(
\widehat{\mathrm{FN}}_1,
\widehat{\mathrm{FN}}_{\mathrm{seq}}
\mid \mathcal S_1
\right)
\right] \\
&\quad+
\operatorname{Cov}_1
\left[
\mathbb{E}_2
\left(
\widehat{\mathrm{FN}}_1
\mid \mathcal S_1
\right),
\mathbb{E}_2
\left(
\widehat{\mathrm{FN}}_{\mathrm{seq}}
\mid \mathcal S_1
\right)
\right] \\
&=
0+
\operatorname{Cov}_1
\left(
\widehat{\mathrm{FN}}_1,
\mathrm{FN}
\right)
=0.
\end{aligned}
\]
The first term is zero because \(\widehat{\mathrm{FN}}_1\) is fixed conditional on the realized Phase~1 sample. The second is zero because \(\mathbb{E}_2(\widehat{\mathrm{FN}}_{\mathrm{seq}}\mid\mathcal S_1)=\mathrm{FN}\) is the fixed finite-population total. The composite variance estimator is therefore
\[
\widehat{\operatorname{Var}}
\left\{
\widehat{\mathrm{FN}}_{\mathrm C}
\right\}
=
\lambda^2
\widehat{\operatorname{Var}}
\left\{
\widehat{\mathrm{FN}}_1
\right\}
+
(1-\lambda)^2
\widehat{\operatorname{Var}}
\left\{
\widehat{\mathrm{FN}}_{\mathrm{seq}}
\mid\mathcal S_1
\right\}.
\]
Because
$
\widehat{\mathrm{TN}}_{\mathrm C}
=
N_--\widehat{\mathrm{FN}}_{\mathrm C},
$
we also have
$
\widehat{\operatorname{Var}}
\left\{
\widehat{\mathrm{TN}}_{\mathrm C}
\right\}
=
\widehat{\operatorname{Var}}
\left\{
\widehat{\mathrm{FN}}_{\mathrm C}
\right\}.
$

\paragraph{Operating-characteristic estimation and CIs.}

We estimate the operating characteristics by substituting \(\widehat{\mathrm{FN}}_{\mathrm C}\) and \(\widehat{\mathrm{TN}}_{\mathrm C}\) for the corresponding SRS HT totals in Equation~\eqref{eq:operating_characteristic_estimators}. Their variance estimators follow the delta-method mappings in Equation~\eqref{eq:delta_method_variances}, using \(\widehat{\operatorname{Var}}(\widehat{\mathrm{FN}}_{\mathrm C})\) and \(\widehat{\operatorname{Var}}(\widehat{\mathrm{TN}}_{\mathrm C})\) derived above. For NPV, sensitivity, specificity, and accuracy, we apply the logit-scale standard-error and CI constructions in Equations~\eqref{eq:logit_se} and \eqref{eq:logit_ci}, respectively. PPV is fully observed within the finite algorithm-positive group. When a CI is reported for PPV under a superpopulation interpretation, we use a Wilson score CI based on \(a\) TPs among \(a+b\) reviewed algorithm-positive patients.

When adaptive allocation is not triggered and no FNs are observed among the final \(m\) algorithm-negative reviews, the usual plug-in variance is zero. In this boundary case, the final negative sample is an SRS sample, which allows an exact finite-population upper confidence bound for the total number of FNs. Define
\[
K_U
=
\max
\left\{
K:
\frac{
\binom{N_--K}{m}
}{
\binom{N_-}{m}
}
\geq
\frac{\alpha}{2}
\right\}.
\]
The resulting confidence set for the finite-population FN total is
$
[0,K_U].
$
We transform this interval monotonically to obtain CIs for the operating characteristics:
\[
\begin{aligned}
\mathrm{Sens} &\in \left[\frac{a}{a+K_U},1\right],
& \mathrm{NPV} &\in \left[\frac{N_--K_U}{N_-},1\right], \\
\mathrm{Spec} &\in \left[\frac{N_--K_U}{N_--K_U+b},\frac{N_-}{N_-+b}\right],
& \mathrm{Acc} &\in \left[\frac{a+N_--K_U}{N},\frac{a+N_-}{N}\right].
\end{aligned}
\]

The adaptive procedure changes how algorithm-negative patients are sampled but not the finite-population estimands. The risk score is used only to define sampling strata and allocate chart reviews; it is not used to impute unreviewed outcomes. Design validity requires prespecification of the strata, total review budget, pilot size, triggering threshold, and allocation constraints. It also requires every patient eligible for Phase~2 sampling to have a known, positive inclusion probability under the realized design.

\section{Simulation}
\subsection{Simulation setup}
We conducted a simulation study to evaluate the finite-sample performance of the proposed validation framework. We quantified finite-sample bias for the design-weighted estimators of sensitivity, specificity, NPV, PPV, and accuracy, and compared them with naive complete-case estimators that ignore the negative-sampling design. We also examined how the adaptive design behaved as FNs became more concentrated across risk strata, compared its RMSE and CI widths with those of SRS under the same chart-review budget, and assessed empirical CI coverage. Each of the 2000 Monte Carlo (MC) replicates first generated a complete finite cohort, with its full set of \(D_i\) values defining the finite-population truth. We then mimicked partial verification by making \(D_i\) available for all algorithm-positive patients and only for algorithm-negative patients selected under the SRS or adaptive two-phase design. Estimates obtained from the partially verified data were compared with the complete-cohort truth. A secondary sensitivity analysis allowed Phase~2 verification to depend weakly on the unobserved outcome within risk strata. All simulations were implemented in \textsf{R 4.5.1}.

For each MC replicate, we generated a finite cohort of $N=3{,}000$ patients, similar to the breast cancer application cohort in Section 5. We generated a continuous covariate
$
X_i \sim N(0,1)
$
and an observed pre-chart-review risk score
$
P_i=\operatorname{logit}^{-1}(\alpha_0+\alpha_1X_i),
$
where $\alpha_1=1.0$ and $\alpha_0$ were selected so that mean risk was 0.30. The score \(P_i\) is observed before chart review and is used only to define risk strata. We generated the true disease status as
\[
D_i\sim\operatorname{Bernoulli}(p_i),
\qquad
p_i=\operatorname{logit}^{-1}(\beta_0+\beta_1X_i),
\]
where $\beta_1=0.8$ and $\beta_0$ was calibrated within each scenario to produce target disease prevalences of 5\%, 10\%, or 20\%. The observed score \(P_i\) defines the sampling strata, whereas \(p_i\) generates disease status. The algorithm classification was generated as
\[
T_i\sim\operatorname{Bernoulli}(q_i),
\qquad
q_i=\operatorname{logit}^{-1}
\left(\gamma_0+\gamma_1D_i+\gamma_3D_iX_i\right).
\]
Here, $q_i$ is a latent probability that controls the simulated algorithm's operating characteristics. This conditional model generates algorithm classifications for the simulation; disease status remains unobserved for unreviewed patients during validation. The parameters $\gamma_0$ and $\gamma_1$ were calibrated to produce two algorithm-performance profiles: high sensitivity with moderate specificity $(0.90,0.85)$ and moderate sensitivity with high specificity $(0.80,0.95)$. We set $\gamma_3$ to 0, $-0.6$, or $-1.2$ to represent no additional, weak, or strong heterogeneity in sensitivity across $X_i$. Because \(P_i\) increases with \(X_i\), when $\gamma_3<0$, cases with larger $X_i$ values are both assigned to higher-risk strata and more likely to be missed by the algorithm. This concentrates FNs in higher-risk strata and creates the risk signal that adaptive sampling is intended to exploit. Algorithm-negative patients were divided into five strata defined by quintiles of $P_i$. We generated the complete data $(D_i,T_i,P_i)$ in each replicate.

We compared two strategies for verifying algorithm-negative patients $(T_i=0)$ while fully verifying all algorithm-positive patients $(T_i=1)$. Under SRS, we sampled $m$ algorithm-negative patients without replacement. Under the adaptive two-phase design, an initial SRS pilot sample of $m_1$ algorithm-negative patients was used to estimate stratum-specific FN rates and calculate the projected sensitivity-variance ratio $\widehat R(m)$. For a paired comparison, SRS and adaptive designs shared the same Phase~1 pilot sample. The SRS design selected the remaining $m_2=m-m_1$ patients by SRS, whereas the adaptive design used the pilot data to determine the Phase~2 allocation. We prespecified \(\rho=1.10\), requiring a projected variance reduction of more than approximately \(9.1\%\) before triggering adaptive allocation. If $\widehat R(m)>1.10$, the remaining $m_2$ charts were allocated across the five risk strata using Neyman allocation. Otherwise, the remaining charts were selected by SRS. This paired construction held the generated cohort, Phase~1 sample, and total chart-review budget fixed across the two designs.

The primary simulation grid varied disease prevalence (5\%, 10\%, or 20\%), sensitivity heterogeneity (none, weak, or strong), algorithm-performance profile (high sensitivity/moderate specificity or moderate sensitivity/high specificity), total negative chart-review size $m$ (150, 275, or 400), and pilot size $m_1$ (50, 100, or 150), yielding 162 scenarios under the assumed probability-sampling design. When $m_1=m$, no Phase~2 sample remained, so the adaptive and SRS designs coincided by construction. As a stress test of departure from the assumed within-stratum probability-sampling design, we repeated the 162 scenarios under an outcome-dependent Phase~2 verification mechanism with an odds ratio of 1.2 for verification of patients with $D_i=1$ versus $D_i=0$ within risk strata, yielding 324 scenarios in total. We conducted 2,000 MC replicates per scenario. In each replicate, the full generated cohort provided the finite-population truth for evaluating bias, RMSE, empirical CI coverage, and average CI width. We also recorded the probability that the adaptive rule triggered Neyman allocation.

\subsection{Results}
Ignoring the unequal verification probabilities of algorithm-positive and algorithm-negative patients produced substantial bias in estimated operating characteristics. Across the 162 primary scenarios, the naive complete-case estimator overestimated sensitivity by 0.129 and underestimated specificity and accuracy by 0.381 and 0.216, respectively. Its bias for NPV was smaller ($-0.002$), and PPV was unaffected because all algorithm-positive patients were reviewed. In contrast, both design-based procedures produced estimates close to the finite-population operating characteristics, showing that accounting for the negative-sampling design removed the substantial bias of the complete-case analysis.

We first assess validity and CI calibration across the full primary simulation grid. Table~\ref{tab:simulation_summary} summarizes MC performance of the adaptive composite and design-based SRS estimators. Both design-based approaches had negligible bias for specificity, NPV, and accuracy. Sensitivity had a small positive bias of approximately 0.005 under both designs, as expected from finite-sample ratio bias in a nonlinear plug-in estimator. Design unbiasedness applies to the estimated confusion-matrix totals, not exactly to every transformed operating characteristic. Sensitivity was the least precise metric, but empirical coverage for metrics affected by negative subsampling remained close to the nominal 95\% level, ranging from 94.1\% to 96.1\%. PPV had no sampling error because the algorithm-positive group was fully reviewed. Averaged across scenarios, adaptive sampling had modestly lower RMSE and mean CI width than SRS.
\begin{table}[!t]
    \centering
    \caption{Simulation performance of adaptive composite and SRS estimators. Results are averaged across 162 primary probability-sampling scenarios. Empirical SD is calculated across MC replicates. Coverage is the empirical coverage of nominal 95\% CIs. Values are reported to three significant digits, except coverage, which is reported as a percentage to one decimal place. Dashes indicate sampling-performance quantities that do not apply because all algorithm-positive patients are reviewed.}
    \label{tab:simulation_summary}
    \small
    \resizebox{\textwidth}{!}{%
    \begin{tabular}{llrrrrr}
        \toprule
        Design & Metric & Signed bias & Empirical SD & RMSE & 95\% CI coverage & Mean CI width \\
        \midrule
        Adaptive & Sensitivity & 0.00462 & 0.0562 & 0.0564 & 95.9\% & 0.225 \\
        Adaptive & Specificity & $-5.66\times10^{-6}$ & 0.000692 & 0.000692 & 94.1\% & 0.00269 \\
        Adaptive & PPV & --- & --- & --- & --- & --- \\
        Adaptive & NPV & $-3.14\times10^{-7}$ & 0.00819 & 0.00819 & 96.0\% & 0.0362 \\
        Adaptive & Accuracy & $6.47\times10^{-7}$ & 0.00660 & 0.00660 & 94.4\% & 0.0258 \\
        \addlinespace
        SRS & Sensitivity & 0.00464 & 0.0569 & 0.0571 & 96.0\% & 0.228 \\
        SRS & Specificity & $-6.75\times10^{-6}$ & 0.000705 & 0.000705 & 94.1\% & 0.00274 \\
        SRS & PPV & --- & --- & --- & --- & --- \\
        SRS & NPV & $-1.10\times10^{-5}$ & 0.00835 & 0.00835 & 96.1\% & 0.0370 \\
        SRS & Accuracy & $-7.96\times10^{-6}$ & 0.00673 & 0.00673 & 94.4\% & 0.0263 \\
        \bottomrule
    \end{tabular}%
    }
\end{table}

\begin{figure}[p]
    \centering
    \includegraphics[width=0.90\textwidth,height=0.78\textheight,keepaspectratio]{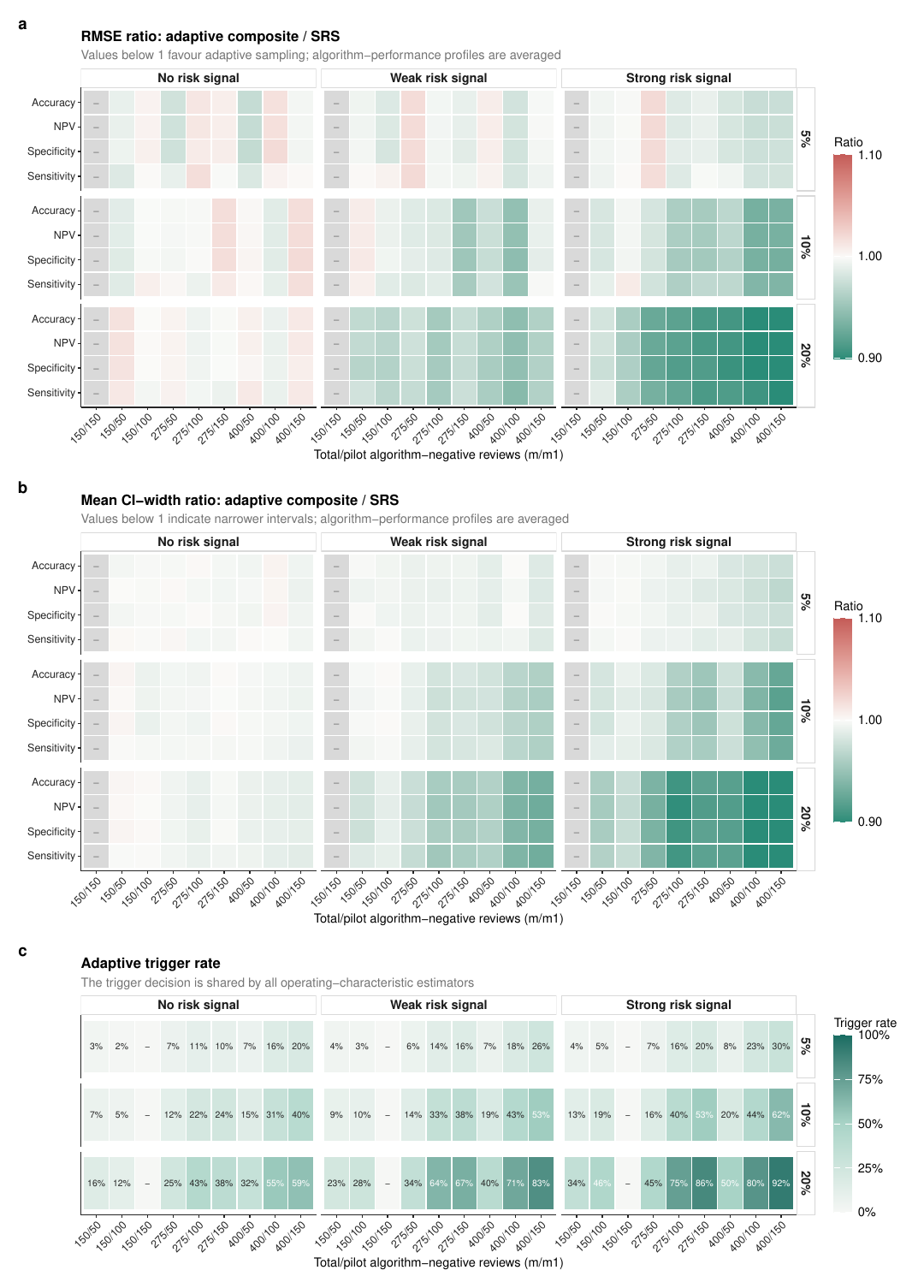}
    \caption{Scenario-specific performance of the adaptive composite estimator relative to design-based SRS under the primary probability-sampling simulations. (\textbf{a}) RMSE ratio and (\textbf{b}) mean 95\% CI-width ratio, calculated as adaptive composite divided by SRS; values below 1 favor adaptive sampling. (\textbf{c}) Probability that the prespecified decision rule triggered adaptive Neyman allocation. Columns indicate the strength of the risk signal, rows indicate disease prevalence, and algorithm-performance profiles are averaged. Cells marked with a dash correspond to $m=m_1$, for which no Phase~2 sample remains and the two designs coincide by construction.}
    \label{fig:simulation_adaptive_performance}
\end{figure}

We next examine when adaptive sampling improves efficiency relative to SRS. Figure~\ref{fig:simulation_adaptive_performance} shows how its relative performance varied across the simulation grid. The ratios in panels a and b compare adaptive composite estimation with design-based SRS; values below 1 favor adaptive sampling. When the risk score contained no information about FN heterogeneity, adaptive sampling performed similarly to SRS. As the risk signal strengthened, adaptive sampling became more efficient, with modest average reductions in RMSE and CI width under a weak signal and reductions of approximately 4\% under a strong signal. The largest gains occurred in scenarios with both strong FN heterogeneity and higher disease prevalence. The trigger rate in Figure~\ref{fig:simulation_adaptive_performance}c increased with disease prevalence, risk-signal strength, and the chart-review budget available after the pilot. The rule sometimes triggered when the risk signal was uninformative, but RMSE and CI-width ratios remained close to 1 in these scenarios. Appendix Figure~\ref{fig:outcome_dependent_verification} reports results from the secondary outcome-dependent verification stress test. Under this deliberate departure from the assumed design, outcome-dependent verification induced additional negative bias and reduced CI coverage for sensitivity and NPV under both designs.

\section{Application to Real-world Data}
Breast cancer recurrence drives subsequent treatment decisions, survival, and healthcare utilization. It can be difficult to detect reliably from structured EHR data because recurrence is not captured as a single standardized feature; instead, it must be inferred from patterns of diagnoses, surgical procedures, and treatments over time. Ritzwoller et al. developed and validated an EHR-based recurrence-detection framework that incorporates clinical indicators, including diagnosis and procedure codes, into a multivariable logistic-regression algorithm called RECUR. RECUR predicts binary recurrence status and timing using a decision threshold selected to optimize performance objectives. \cite{Ritzwoller2018} We adapted this algorithm to Duke University Health System (DUHS) EHR and tumor-registry data by translating each indicator to DUHS-specific data structures and coding systems, applying the algorithm, and using the accuracy-maximizing threshold for recurrence detection. We then evaluated performance in the DUHS cohort using the proposed two-phase validation design with chart review. The design yields design-unbiased estimates of the TP, FP, FN, and TN totals, from which operating characteristics are estimated under partial verification.

As chart-reviewed recurrence status was available for all patients in the complete DUHS cohort, this application also supports a repeated-sampling evaluation of the validation design. We first describe the cohort, outcome definition, and RECUR implementation. We then treat complete-cohort operating characteristics as the truth and compare SRS with the adaptive procedure across repeated samples from this fixed cohort.

\subsection{Data source and patient cohort}
We applied the validation design to a retrospective cohort of breast cancer patients identified from the DUHS tumor registry. The cohort included patients diagnosed between January 1, 2015, and December 31, 2023. To align with the target population for the recurrence algorithm, patients were included if they (1) had Stage I--III incident breast cancer, (2) were aged $\ge$ 21 years at first diagnosis, (3) underwent definitive surgery for incident breast cancer, and (4) remained alive with documented follow-up for at least 30 days after surgery. To restrict the cohort to primary incident cases, we excluded patients with a prior history of malignancy or records of multiple primary tumors. The final eligible cohort contained 3,336 patients. We extracted the diagnosis codes, procedure codes, medication records, and other clinical data required for RECUR from the DUHS EHR and linked them to tumor-registry entries.

Recurrence was originally recorded with event timing, whereas the proposed validation framework addresses binary operating characteristics. We evaluated binary recurrence-detection performance, consistent with the detection component of the original RECUR study, using prespecified 3-, 4-, and 5-year endpoints after the index diagnosis date to define a common assessment horizon.\citep{Ritzwoller2018} For each endpoint \(y\in\{3,4,5\}\), we defined \(D_i^{(y)}=1\) when chart review confirmed recurrence on or before \(y\) years after diagnosis and \(D_i^{(y)}=0\) when chart review documented no recurrence through that endpoint. Patients without documented recurrence were classified as recurrence-free at endpoint \(y\) only if their chart contained follow-up extending at least \(y\) years after diagnosis. Patients without recurrence but with insufficient follow-up, and records with an unavailable diagnosis date or unusable recurrence date, were not evaluable for that endpoint. These evaluability criteria ensured that recurrence status could be determined over the same prespecified time horizon for all patients included in each endpoint analysis. The 3-, 4-, and 5-year analytic cohorts included 2,503, 2,051, and 1,649 patients, respectively. The corresponding algorithm classification \(T_i^{(y)}\) indicated whether RECUR predicted recurrence by the same endpoint. 

To establish the gold-standard outcome, M. B., L. Z., J. A., and S. M. manually reviewed 3,336 eligible patient records using a prespecified abstraction protocol. Reviewers ascertained recurrence from clinic notes in which a medical oncologist, surgeon, or radiation oncologist documented recurrence. For patients with recurrence, the recurrence date was defined as the date of a biopsy confirming recurrence; when no biopsy was performed, the date of imaging showing likely recurrence was used. These chart-reviewed outcomes provided the gold standard for all patients in each endpoint-specific analytic cohort. We used the resulting complete-cohort operating characteristics as the finite-population truth in the repeated-sampling evaluation of the SRS and adaptive validation designs.

\subsection{Algorithm implementation and results}
Details of RECUR development and evaluation are reported by Ritzwoller et al. and Hassett et al.\cite{Ritzwoller2018, RECUR2017} The algorithm uses seven predictors: secondary malignancy excluding lymph-node involvement, radiation therapy, mastectomy, high-cost imaging, chemotherapy, hospice, and observation encounters. Appendix Table~\ref{tab:administrative_codes} summarizes the DUHS EHR data used to construct these predictors, including International Classification of Diseases, Tenth Revision (ICD-10) and Ninth Revision (ICD-9) diagnosis codes; Current Procedural Terminology (CPT), Healthcare Common Procedure Coding System (HCPCS), and ICD-10-Procedural Coding System (PCS) procedure codes; Diagnosis Related Group (DRG) codes; generic drug names; place-of-service codes; and revenue codes. The table is directly adapted from the supplementary material of Ritzwoller et al. The logistic-regression probability cutoff was 34.1\% to maximize accuracy. \cite{Ritzwoller2018} The original algorithm was implemented in \textsf{SAS}, and we translated the code to \textsf{R}. All analyses were conducted in \textsf{R 4.5.1}. In the repeated-sampling evaluation, the continuous RECUR predicted probability served as the pre-review risk score \(P_i\).

Table~\ref{tab:baseline-characteristics} summarizes patient characteristics and recurrence-detection results for the application cohort. Mean age at diagnosis was 64.9 years. Most patients were White (67.2\%) or Black or African American (23.8\%). Stage~I disease accounted for 65.9\% of the cohort, followed by Stage~II (25.4\%) and Stage~III (8.7\%). Among patients in the endpoint-specific analytic cohorts, documented recurrence increased from 108 of 2,503 patients (4.31\%) within 3 years to 156 of 1,649 patients (9.46\%) within 5 years, although endpoint availability decreased with longer follow-up.

For context, the 34.1\% accuracy-maximizing threshold used in our implementation was also the accuracy-maximizing threshold in the original RECUR study. In the original external-validation cohort, Ritzwoller et al. reported sensitivity of 74.7\%, specificity of 99.1\%, PPV of 86.8\%, and NPV of 97.9\% at this threshold.\cite{Ritzwoller2018} In the DUHS cohort, the corresponding finite-population estimates varied across the 3-, 4-, and 5-year endpoints (Table~\ref{tab:real_data_validation}). Unlike the original study, which evaluated recurrence over variable follow-up in a different health system, our analysis evaluated prespecified fixed-endpoint recurrence outcomes in the DUHS cohort. 

\begin{table}[!t]
\centering
\caption{Patient characteristics and endpoint-specific recurrence outcomes in the DUHS cohort.}
\label{tab:baseline-characteristics}
\footnotesize
\renewcommand{\arraystretch}{0.90}
\begin{tabular}{p{0.73\textwidth}>{\raggedleft\arraybackslash}p{0.20\textwidth}}
\toprule
Characteristic & Overall \\
\midrule
\textbf{Patient characteristics} & \textbf{N = 3{,}336} \\
Age, mean (SD) & 64.9 (13.2) \\
\addlinespace
\textbf{Race, n (\%)} & \\
\quad Black or African American & 793 (23.8\%) \\
\quad White & 2{,}242 (67.2\%) \\
\quad Other or unknown & 301 (9.0\%) \\
\addlinespace
\textbf{Sex, n (\%)} & \\
\quad Female & 3{,}323 (99.6\%) \\
\quad Male & 12 (0.4\%) \\
\quad Other & 1 ($<0.1$\%) \\
\addlinespace
\textbf{Ethnicity, n (\%)} & \\
\quad Hispanic & 107 (3.2\%) \\
\quad Not Hispanic & 3{,}062 (91.8\%) \\
\quad Unknown & 167 (5.0\%) \\
\addlinespace
\textbf{Cancer stage, n (\%)} & \\
\quad I & 2{,}200 (65.9\%) \\
\quad II & 847 (25.4\%) \\
\quad III & 289 (8.7\%) \\
\midrule
\textbf{Recurrence type among patients with documented recurrence, n (\%)}\textsuperscript{a} & \textbf{n = 216} \\
\quad Distant, including combined or sequential recurrences & 125 (57.9\%) \\
\quad Local or regional only & 88 (40.7\%) \\
\quad Type unavailable & 3 (1.4\%) \\
\midrule
\textbf{Endpoint-specific recurrence: events / analytic patients (\%)}\textsuperscript{b} & \\
\quad Within 3 years & 108 / 2{,}503 (4.31\%) \\
\quad Within 4 years & 136 / 2{,}051 (6.63\%) \\
\quad Within 5 years & 156 / 1{,}649 (9.46\%) \\
\bottomrule
\end{tabular}
\par\smallskip
\begin{minipage}{\textwidth}
\footnotesize
\textsuperscript{a}Recurrence-type percentages use the 216 patients with chart-review-documented recurrence as the denominator, irrespective of the fixed endpoints. Descriptions containing distant recurrence, including ``Distal,'' were grouped as distant.\\[3pt]
\textsuperscript{b}Endpoint-specific results use patients with evaluable chart-reviewed outcomes and available algorithm classifications who met the analytic eligibility criteria.
\end{minipage}
\renewcommand{\arraystretch}{1}
\end{table}

Appendix Table~\ref{tab:clinical-events-by-stage} reports the distributions of the seven clinical-event components used by RECUR, by cancer stage, providing clinical context for the algorithm inputs in the DUHS cohort.

\subsection{Repeated-sampling evaluation of the validation design}
We evaluated the proposed validation framework for recurrence within 3, 4, and 5 years after the index diagnosis date. At each endpoint, the complete-cohort chart-reviewed outcomes defined the finite-population truth. The SRS-based planning procedure selected total algorithm-negative review budgets of $m=125$, 150, and 175 for the 3-, 4-, and 5-year endpoints, respectively. We formed five prespecified risk strata from quintiles of \(P_i\) among algorithm-negative patients, used a reference Phase~1 pilot size of $m_1=75$, and set the adaptive triggering threshold to $\widehat R>1.10$.

For each endpoint, we conducted 500 paired repeated-sampling replicates. In each replicate, all algorithm-positive patients were retained and the same SRS Phase~1 pilot sample was used for both procedures. The SRS procedure completed the remaining $m_2$ reviews by SRS, whereas the adaptive procedure used the pilot data to choose between continued SRS and Neyman allocation. We compared the resulting estimates, RMSE, and CI coverage with the complete-cohort truth. Reported adaptive results summarize the complete prespecified procedure across triggered and non-triggered replicates.

\begin{figure}[!htbp]
    \centering
    \includegraphics[width=0.98\textwidth]{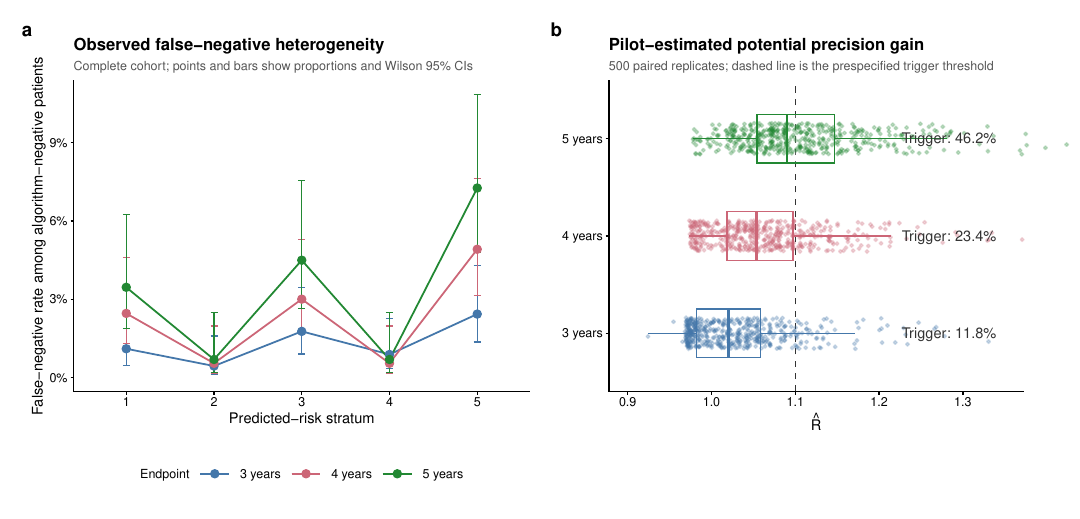}
    \caption{Risk-stratification diagnostics for the real-data application. (\textbf{a}) Complete-cohort FN proportions among algorithm-negative patients across the five predicted-risk strata; bars denote Wilson 95\% CIs. (\textbf{b}) Distribution of the pilot-estimated variance ratio $\widehat R$ across 500 paired repeated-sampling replicates with $m_1=75$. Boxplots show medians and interquartile ranges, points denote individual replicates, and the dashed line marks the prespecified adaptive trigger threshold of 1.10. Labels report the proportion of replicates that triggered Neyman allocation.}
    \label{fig:real_data_risk_diagnostics}
\end{figure}

Table~\ref{tab:real_data_validation} summarizes the numerical results. Finite-population recurrence prevalence increased from 4.31\% at 3 years to 9.46\% at 5 years, while the mean pilot-estimated variance ratio $\widehat R$ increased from 1.03 to 1.11. The adaptive rule triggered Neyman allocation in 11.8\%, 23.4\%, and 46.2\% of replicates for the 3-, 4-, and 5-year endpoints, respectively. Figure~\ref{fig:real_data_risk_diagnostics} provides two diagnostics for interpreting these results: panel a shows whether complete-cohort FN rates differed across risk strata, and panel b shows whether the pilot data identified sufficient projected gain to trigger adaptive allocation. Because the decision is made using replicate-specific \(\widehat R\), its mean does not itself represent the trigger probability. Mean estimates of specificity, PPV, NPV, and accuracy were close to their finite-population values under both designs. PPV was identical across replicates. Sensitivity showed modest positive finite-sample bias under both designs at all three endpoints.

% Generated by analysis_scripts/build_real_data_repeated_sampling_table.R
\begin{table}[!t]
\centering
\caption{Repeated-sampling performance in the real-data application. Results use $m_1=75$ and 500 paired replicates. Percentages are reported to one decimal place; other noninteger values are reported to three significant digits. Dashes indicate sampling-performance quantities that do not apply because all algorithm-positive patients are reviewed.}
\label{tab:real_data_validation}
\scriptsize
\resizebox{\textwidth}{!}{%
\begin{tabular}{lrrrlrccrrrr}
\toprule
 & & & & & & \multicolumn{2}{c}{Mean estimate (95\% empirical CI)} & \multicolumn{2}{c}{RMSE} & \multicolumn{2}{c}{95\% CI coverage} \\
\cmidrule(lr){7-8} \cmidrule(lr){9-10} \cmidrule(lr){11-12}
Endpoint & Prevalence & $\widehat R$ & $m$ & Metric & Truth & Adaptive & SRS & Adaptive & SRS & Adaptive & SRS \\
\midrule
3 years & 4.3\% & 1.03 & 125 & Sensitivity & 0.630 & 0.656 (0.446, 0.872) & 0.655 (0.452, 0.872) & 0.137 & 0.136 & 97.6\% & 97.6\% \\
 &  &  &  & Specificity & 0.972 & 0.972 (0.971, 0.972) & 0.972 (0.971, 0.972) & 0.000301 & 0.000274 & 99.8\% & 100.0\% \\
 &  &  &  & PPV & 0.500 & 0.500 & 0.500 & --- & --- & --- & --- \\
 &  &  &  & NPV & 0.983 & 0.983 (0.964, 0.996) & 0.983 (0.965, 0.996) & 0.011 & 0.010 & 97.6\% & 97.6\% \\
 &  &  &  & Accuracy & 0.957 & 0.957 (0.939, 0.969) & 0.957 (0.940, 0.969) & 0.010 & 0.009 & 99.2\% & 99.4\% \\
\addlinespace
4 years & 6.6\% & 1.06 & 150 & Sensitivity & 0.647 & 0.663 (0.491, 0.936) & 0.657 (0.491, 0.936) & 0.108 & 0.111 & 97.8\% & 97.6\% \\
 &  &  &  & Specificity & 0.958 & 0.958 (0.957, 0.959) & 0.958 (0.957, 0.959) & 0.000476 & 0.000508 & 89.8\% & 89.6\% \\
 &  &  &  & PPV & 0.524 & 0.524 & 0.524 & --- & --- & --- & --- \\
 &  &  &  & NPV & 0.975 & 0.974 (0.951, 0.997) & 0.974 (0.951, 0.997) & 0.012 & 0.012 & 98.0\% & 97.6\% \\
 &  &  &  & Accuracy & 0.938 & 0.938 (0.916, 0.958) & 0.937 (0.916, 0.958) & 0.011 & 0.011 & 88.8\% & 87.6\% \\
\addlinespace
5 years & 9.5\% & 1.11 & 175 & Sensitivity & 0.679 & 0.691 (0.542, 0.851) & 0.694 (0.556, 0.851) & 0.082 & 0.080 & 96.2\% & 97.0\% \\
 &  &  &  & Specificity & 0.944 & 0.944 (0.942, 0.945) & 0.944 (0.942, 0.945) & 0.000786 & 0.000681 & 93.6\% & 94.6\% \\
 &  &  &  & PPV & 0.558 & 0.558 & 0.558 & --- & --- & --- & --- \\
 &  &  &  & NPV & 0.966 & 0.966 (0.939, 0.987) & 0.967 (0.942, 0.987) & 0.014 & 0.012 & 96.2\% & 97.0\% \\
 &  &  &  & Accuracy & 0.919 & 0.919 (0.895, 0.938) & 0.919 (0.898, 0.938) & 0.012 & 0.011 & 93.4\% & 94.2\% \\
\bottomrule
\end{tabular}%
}
\end{table}

\begin{figure}[!htbp]
    \centering
    \includegraphics[width=0.98\textwidth,height=0.86\textheight,keepaspectratio]{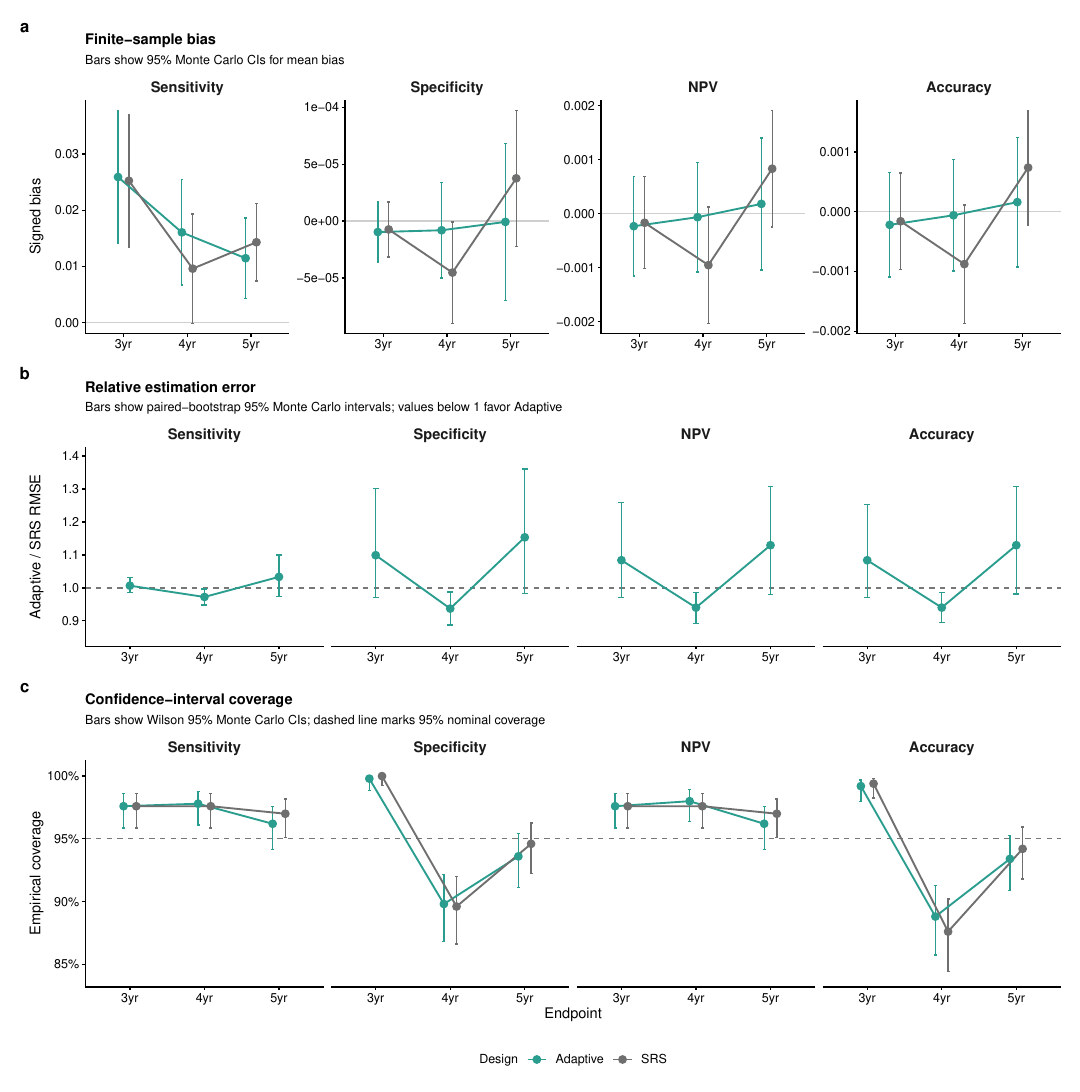}
    \caption{Real-data repeated-sampling performance at the reference pilot size $m_1=75$, based on 500 paired replicates per endpoint. (\textbf{a}) Signed bias relative to the finite-population truth; error bars are 95\% MC CIs for mean bias. (\textbf{b}) Ratio of adaptive to SRS RMSE; values below 1 favor the adaptive procedure, and error bars are paired-bootstrap 95\% CIs. (\textbf{c}) Empirical coverage of replicate-specific 95\% CIs; error bars are Wilson 95\% MC CIs and the dashed line marks 95\% nominal coverage. PPV is not shown because its estimate was unaffected by algorithm-negative sampling.}
    \label{fig:real_data_repeated_sampling}
\end{figure}

Figure~\ref{fig:real_data_repeated_sampling} then compares point-estimation and interval performance for the four operating characteristics affected by algorithm-negative subsampling. MC CIs for signed bias included zero for specificity, NPV, and accuracy, whereas sensitivity retained the small positive ratio bias described above (Figure~\ref{fig:real_data_repeated_sampling}a). Adaptive-to-SRS RMSE ratios were below 1 for all four metrics at the 4-year endpoint but exceeded 1 for most metrics at the 3- and 5-year endpoints (Figure~\ref{fig:real_data_repeated_sampling}b). The paired-bootstrap CIs generally included 1 at 3 and 5 years. Thus, adaptive sampling did not consistently reduce RMSE.

Empirical coverage for sensitivity and NPV was above 95\% across endpoints and designs (Figure~\ref{fig:real_data_repeated_sampling}c). Specificity and accuracy showed lower coverage at the 4-year endpoint and slightly below-nominal coverage at the 5-year endpoint. Coverage patterns were similar for adaptive and SRS procedures. PPV was omitted from Figure~\ref{fig:real_data_repeated_sampling} because its estimate did not depend on algorithm-negative sampling.

\clearpage
\section{Discussion}
Our findings support a finite-population, design-based approach to validation when all algorithm-positive patients can be reviewed but only a sample of algorithm-negative patients is feasible. Ignoring the unequal verification probabilities of algorithm-positive and algorithm-negative patients produced substantial bias in operating-characteristic estimates from the chart-reviewed sample, whereas accounting for the negative-sampling design removed this bias in the simulations. Adaptive sampling improved efficiency relative to SRS when the risk score meaningfully identified FN heterogeneity, but it did not guarantee gains over SRS. In the recurrence application, adaptive and SRS procedures had similar signed bias and broadly similar coverage despite an inconsistent adaptive efficiency advantage, indicating that the adaptive extension preserved validity of the performance estimates.

These results distinguish validity from efficiency. The composite estimator and its variance estimator were introduced to stabilize estimation and inference under the two-phase design. McIsaac and Cook described a trade-off between estimating allocation parameters precisely from an internal pilot and retaining enough observations for subsequent adaptive sampling.\citep{McIsaacCook2015} Marks-Anglin et al. similarly noted that surrogate-assisted sampling may offer less benefit when a small pilot provides limited information about a rare outcome.\citep{MarksAnglin2025} In our recurrence application, even ideal Neyman allocation based on complete-cohort information offered only modest projected gains, with SRS-to-Neyman variance ratios of 1.08, 1.19, and 1.24 for the 3-, 4-, and 5-year endpoints. This limited potential gain, together with uncertainty in pilot-estimated FN rates, helps explain why the practical adaptive procedure did not consistently improve on SRS.

Accordingly, adaptive allocation should be viewed as an optional extension of SRS, with its use guided by the estimation target and the information available for allocation. Chen and Lumley showed that multiwave sampling can approximate optimal designs for regression-parameter estimation when informative design inputs are available.\citep{ChenLumley2020} In sequential chart validation, Hahn et al. also found that random sampling often performed similarly to Neyman sampling and emphasized its practical simplicity.\citep{HahnAdaptive} Zou et al. use auxiliary information differently, estimating true-outcome probabilities from chart-reviewed data and incorporating them into performance estimation.\citep{Zou2026} Our framework addresses a related but distinct decision: how to allocate a prespecified algorithm-negative review budget to estimate finite-cohort operating characteristics. SRS-based precision targets determine the review budget, and the Phase~1 pilot guides allocation of the remaining reviews. When the projected gain does not exceed the prespecified threshold, the procedure continues with SRS and the composite estimator reduces to the ordinary SRS estimator.

For study planning and reporting, investigators can select the total number of algorithm-negative reviews using SRS-based CI-width targets, then use the Phase~1 pilot to determine whether the remaining reviews should be allocated differentially across risk strata. In our framework, the risk score guides chart selection, while performance estimation uses reviewed outcomes and known sampling probabilities. Investigators should report the realized sampling design, review counts by stratum, inclusion probabilities, and design-based estimates. Gold-standard outcomes were available for every patient in each endpoint-specific analytic cohort, allowing repeated-sampling evaluation of trigger rates, RMSE ratios, and CI coverage.

The evaluable cohort became smaller at longer endpoints because patients without documented recurrence required follow-up through the corresponding horizon. Consequently, the endpoint-specific estimates describe different subsets of the original cohort. Although recurrence dates and follow-up were used to determine endpoint-specific status, the binary analysis did not model recurrence timing or incorporate information from patients censored before the endpoint without documented recurrence. A time-to-event extension would allow these features to be incorporated. The real-world application involved one health system, one algorithm, and a finite cohort with relatively modest FN heterogeneity. The efficiency results may therefore not generalize to phenotypes or algorithms with stronger or more reliably predicted heterogeneity. 

Future work should develop trigger rules that account for uncertainty in the pilot-based variance ratio, for example through shrinkage or model-assisted estimation of stratum-specific FN rates. Further evaluation in external EHR cohorts and phenotypes with stronger risk-outcome heterogeneity would clarify when adaptive allocation yields meaningful reductions in review burden or CI width.

\section{Acknowledgments}
This work was funded by a pilot grant from the Duke Cancer Institute's (DCI) Center for Onco-Primary Care, which was made possible by NCI Cancer Center Support Grant (CCSG) P30CA014236. The content is solely the responsibility of the authors and does not necessarily represent the official views of the National Institutes of Health.

Generative AI tools (ChatGPT and Codex, OpenAI; accessed June--August 2026) were used to assist with restructuring and language editing of portions of the manuscript to improve clarity, flow, and cohesion. All suggested text was critically reviewed, revised where appropriate, and approved by the authors. The authors take full responsibility for the scientific content, accuracy, and conclusions of this work.

\section{Data Availability Statement}
The Duke University Health System EHR data and chart-review data used in this study are not publicly available because they contain protected health information and are subject to institutional data-use restrictions. Access to these data requires approval from DUHS and the relevant institutional review bodies. De-identified simulation code, a synthetic example dataset, and aggregate source data sufficient to reproduce the reported tables and figures will be made available in a public repository upon publication.

Generative AI tools (ChatGPT and Codex, OpenAI; accessed February--August 2026) were used to assist with drafting and debugging R code. All code and analyses were independently reviewed and approved by the authors. No patient-level data were provided to the AI tools, and the tools did not generate study data or make final analytic or interpretive decisions.

\bibliographystyle{unsrtnat}
\bibliography{refs}

\clearpage
\appendix
\renewcommand{\thetable}{A\arabic{table}}
\renewcommand{\thefigure}{A\arabic{figure}}
\setcounter{table}{0}
\setcounter{figure}{0}
\renewcommand{\theHtable}{appendix.\arabic{table}}
\renewcommand{\theHfigure}{appendix.\arabic{figure}}
\section{Appendix}
\footnotesize
\begin{longtable}{@{}>{\raggedright\arraybackslash}p{0.22\textwidth}>{\raggedright\arraybackslash}p{0.15\textwidth}>{\raggedright\arraybackslash}p{0.55\textwidth}@{}}
\caption{Administrative data codes used as inputs to the RECUR algorithm. Adapted from the supplementary material of Ritzwoller et al.\citep{Ritzwoller2018}}\label{tab:administrative_codes}\\
\toprule
\textbf{Algorithm Category} & \textbf{Code Type} & \textbf{Code} \\
\midrule
\endfirsthead
\multicolumn{3}{@{}l}{\textit{Appendix Table \thetable{} continued}}\\
\toprule
\textbf{Algorithm Category} & \textbf{Code Type} & \textbf{Code} \\
\midrule
\endhead
\midrule
\multicolumn{3}{r@{}}{\textit{Continued on next page}}\\
\endfoot
\bottomrule
\endlastfoot
Secondary Malignancy Excluding Lymph Node Involvement & ICD9\newline{}ICD10 & 197.x, 198.x, excluding 198.81\newline{}C78.x, C79.x, excluding C79.81 \\
Radiation Therapy & CPT4 & 63620-63621, 76950, 76960, 76965, 77261-77263, 77280, 77285, 77290, 77295, 77299, 77300,77301, 77305, 77310, 77315, 77321, 77326,-77328, 77331-77334, 77336, 77338, 77370-77373, 77399, 77401-77423,  77425, 77427, 77430-77432, 77435, 77763, 77776-77778, 77781-77790, 77799, 79000, 79001, 79005, 79020, 79030, 79035, 79100, 79101, 79200, 79300, 79400, 79403, 79420, 79440, 79445, 79900, 79999 \\
 & DRG & 409 \\
 & HCPC & A4650, A9517, A9518, A9523, A9527, A9530, A9534, A9543, A9545, A9563, A9564, A9600, A9603, A9605, A9699, C1064, C1065, C1081, C1083, C1188, C1348, C1715-C1720, C1728, C2616, C2633-C2635, C2637-C2643, C2698, C2699, C9401, C9402, C9405, C9725, C9726, C9728, G0173, G0174, G0178, G0242, G0243, G0251, G0256, G0261, G0273, G0274, G0338-G0340, G8379, Q3001, S2270, S8049 \\
 & ICD9\newline{}ICD10 & V58.0, V66.1, V67.1\newline{}Z51.0, Z51.89, Z08, Z09 \\
Mastectomy & CPT4 & 19180, 19182, 19184-19187, 19200, 19211-19216, 19220, 19224-19229, 19240, 19250-19255, 19303-19307 \\
 & DRG & 257-258 (for claims prior to 31Dec2008)582-583 (for claims post 01Jan2009) \\
 & ICD9-CM\newline{}ICD-10-PCS & 85.33-85.36, 85.40-85.48\newline{}0H0T0JZ, 0H0T3JZ, 0H0U0JZ, 0H0U3JZ, 0HRT0JZ, 0HRT3JZ, 0HRU0JZ, 0HRU3JZ, 0HRV0JZ, 0HRV3JZ, 0HBT0ZZ, 0HBT3ZZ, 0HBU0ZZ, 0HBU3ZZ, 0H0V0JZ, 0H0V3JZ, 0HBV0ZZ, 0HBV3ZZ, 0HTT0ZZ, 0HTU0ZZ, 0HTV0ZZ, 07T50ZZ, 0HTT0ZZ, 07T60ZZ, 0HTU0ZZ, 07T60ZZ, 0HTV0ZZ, 0KTH0ZZ, 0KTJ0ZZ, 07T70ZZ, 07T80ZZ, 07T90ZZ \\
High Cost Imaging & CPT4 & 8703-8704, 8741-8742, 8771, 8801-8802, 8838, 8891-8895, 8897-8898, 9201-9205, 9209, 9211-9219, 70336, 70450-70492, 70496-70498, 70540-70549, 70551-70555, 70559, 71250-71270, 71275, 71550-71552, 71555, 72125-72133, 72141-72159, 72191-72198, 73200-73202, 73206, 73218, 73219-73223, 73225, 73700-73702, 73706, 73718-71723, 71725, 74150-74170, 74175, 74181-74185, 74261-74263, 75552-75564, 75635, 76070-76071, 76093, 76094, 7140, 76380, 76390, 76400, 76497-76498, 77058-77059, 77078-77079, 77084, 78000- 78018, 78020, 78070, 78075, 78099, 78102-78104, 78110-78111, 78120-78122, 78130-78135, 78140, 78160-78162, 78170-78172, 78185, 78190-78191, 78195, 78199, 78201, 78202, 78205, 78206, 78215, 78216, 78220, 78223, 78230-78232, 78258, 78261, 78262, 78264, 78267, 78268, 78270, 78271, 78278, 78282, 78290, 78291, 78299-78315, 78320, 78350, 78351, 78399, 78414, 78428, 78445, 78456--8469, 78472, 78473, 78478, 78480-78483, 78491, 78492, 78494, 78496, 78499, 78580-78588, 78591-78594, 78596, 78599-78610, 78615, 78630, 78635, 78645, 78647, 78650, 78660, 78699-78710, 78715, 78725, 78730, 78740, 78760, 78761, 78799-78806, 78810-78816, 78999, 0042T, 0066T, 0066T, 0067T, 0144T, 0150T, 0151T, 0155T, \\
 & HCPC & C8900-C8920, G0030-G0047, G0125, G0126, G0131, G0132, G0163-G0165, G0210-G0235, G0252-G0254, G0296, G0330, G0331, G0336, S8037, S8042, S8092, S8093 \\
Chemotherapy & ICD9 & V58.1. V58.11, V58.12, V07.3, V07.39\newline{}Z51.11, Z51.12, Z29.8 \\
 & ICD9 CM\newline{}ICD-10-PCS & 0.1, 0.15, 17.7, 99.25, 99.28, 99.29\newline{}3E03303, 3E0330M, 3E04303, 3E0430M, 3E05303, 3E0530M, 3E06303, 3E0630M, 83E03305, 3E04305, XW033B3, XW043B3 \\
 & REV & 331, 332, 335 \\
 & CPT4 & 36640, 61517, 81350, 83520, 96400-96406, 96408-96417, 96420, 96422, 96425, 96440, 96445, 96446, 96450, 96450, 96542, 96545, 96549, 96567, 96570, 96571, 96910, 96912, 203682, 206820, 206929, 209622, 214693, 219583, 219687, 242226, 0519F, 4180F \\
 & HCPC & A9600, A9604, A9605, C1084, C1086, C1166, C1167, C1178, C8953, C8954, C8955, C9004, C9012, C9110, C9131, C9205, C9207, C9213-C9215, C9235, C9257, C9262, C9265, C9284, C9287, C9414-C9429, C9431-C9434, C9437, C9438, C9440, G0355-G0362, G8372-G8374, G9021-G9032, J0178, J0490, J0594, J0894, J1050, J1051, J1094, J1100, J1190, J1457, J1620, J1675, J2323, J2352, J2353, J2354, J3262, J3590, J7150, J7527, J8510, J8520, J8521, J8530, J8540, J8560, J8561, J8562, J8565, J8600, J8610, J8700, J8705, J8999, J9000-J9002, J9010, J9015, J9017, J9019, J9020, J9025, J9027, J9031, J9033, J9035, J9040-J9043, J9045, J9050, J9055, J9060, J9062, J9065, J9070, J9080, J9090-J9098, J9100, J9110, J9120, J9130, J9140, J9150, J9151, J9155, J9160, J9165, J9170, J9171, J9178-J9182, J9185, J9190, J9200-J9202, J9206-J9208, J9211-J9215, J9225, J9226, J9228, J9230, J9245, J9250, J9260, J9261, J9263-J9268, J9270, J9280, J9290, J9291, J9293, J9299, J9300, J9302, J9303-J9307, J9310, J9315, J9320, J9328, J9330, J9340, J9350, J9351, J9354, J9355, J9357, J9360, J9370, J9375, J9380, J9390, J9395, J9400, J9600, J9999, Q0083-Q0085, Q2017, Q2020, Q2024, Q2043, Q2044, Q2046, Q2049, Q2050, Q4079, S0087, S0088, S0115, S0116, S0145, S0146, S0156, S0170, S0172, S0175, S0176, S0178, S0179, S0182, S0187, S2107, S5019, S5020, S9329, S9330, S9331 \\
 & Generic Drug Name & AFATINIB, AFLIBERCEPT, ALDESLEUKIN, ALECTINIB, ALEMTUZUMAB, ALFERON, ALTRETAMINE, ANASTROZOLE, ARSENIC, ASPARAGINASE, AXITINIB, AZACITIDINE, BCG LIVE, BELIMUMAB, BENDAMUSTINE, BETASERON, BEVACIZUMAB, BEXAROTENE, BICALUTAMIDE, BLEOMYCIN, BORTEZOMIB \\
 &  & BOSUTINIB, BRENTUXIMAB, BUSULFAN, CABAZITAXEL, CABOZANTINIB, CAPECITABINE, CARBOPLATIN, CARFILZOMIB, CARMUSTINE, CERITINIB, CETUXIMAB, CHLORAMBUCIL, CISPLATIN, CLADRIBINE, CLOFARABINE, COBIMETINIB, CRIZORINIB, CYCLOPHOSPHAMIDE, CYTARABINE \\
 &  & DACARBAZINE, DACTINOMYCIN, DARATUMUMAB, DASATINIB, DAUNORUBICIN, DECITABINE, DEGARELIX, DENILEUKIN, DIETHYLSTILBESTROL, DINUTUXIMAB, DOCETAXEL, DOXORUBICIN, ELOTUZUMAB, ENZALUTAMIDE, EPIRUBICIN, ERIBULIN, ERLOTINIB, ESTRADURIN, ESTRAMUSTINE \\
 &  & ETOPOSIDE, EVEROLIMUS, EXEMESTANE, FLOXURIDINE, FLUOROURACIL, FLUTAMIDE, FULVESTRANT, GEFITINIB, GEMCITABINE, GEMTUZUMAB, GOSERELIN, HISTRELIN, HYDROXYUREA, IBRUTINIB, IDARUBICIN, IFOSFAMIDE, IFOSFAMIDE/MESNA, IMATINIB, INTERFERON ALFA-2A \\
 &  & INTERFERON ALFA-2B, IPILIMUMAB, IRINOTECAN, IXABEPILONE, IXAZOMIB, KETOCONAZOLE, LAPATINIB, LENALIDOMIDE, LENVATINIB, LETROZOLE, LEUCOVORIN, LOMUSTINE, MECHLORETHAMINE, MEDROXYPROGESTERONE, MEGESTROL, MELPHALAN, MERCAPTOPURINE, MESNA, METHOTREXATE \\
 &  & METHOXSALEN, MITOMYCIN, MITOXANTRONE, NATALIZUMAB, NECITUMUMAB, NELARABINE, NILOTINIB, NIVOLUMAB, NOLVADEX, OBINUTUZUMAB, OCTREOTIDE, OFATUMUMAB, OMACETAXINE, OSIMERTINIB, OXALIPLATIN, PACLITAXEL, PANITUMUMAB, PANOBINOSTAT, PAZOPANIB, PEGASPARGASE \\
 &  & PEMETREXED, PENTOSTATIN, PERTUZUMAB, PLICAMYCIN, POMALIDOMIDE, PONATINIB, PORFIMER, PRALATREXATE, PROCARBAZINE, RADIUM RA 223, RAMUCIRUMAB, REGORAFENIB, RITUXIMAB, ROMIDEPSIN, SAMARIUM, SIPULEUCEL, SORAFENIB, STREPTOZOCIN, STRONTIUM-89, SUNITINIB \\
 &  & TAFINLAR, TALIMOGENE, TAMOXIFEN, TEMOZOLOMIDE, TEMSIROLIMUS, TENIPOSIDE, TESTOLACTONE, THALIDOMIDE, THIOGUANINE, THIOTEPA, TOPOTECAN, TOSITUMOMAB, TRABECTEDIN, TRAMETINIB, TRASTUZUMAB, TRETINOIN, URACIL, VALRUBICIN, VANDETANIB, VEMURAFENIB \\
 &  & VINBLASTINE, VINCRISTINE, VINORELBINE, VISMODEGIB, VORINOSTAT \\
Hospice & Place of Service Code & 34 = Inpatient Hospice \\
 & Revenue Code & 115, 125, 135, 145, 155, 235, 650-653, 655-659 \\
Observation Encounters & CPT & 99217-99220, 99234-99236 \\
 & Revenue Code & 760-762, 769 \\
 & Pre-authorization code & 44317, OM, OO \\
\end{longtable}
\normalsize

\clearpage
\footnotesize
\begin{table}[!t]
\centering
\caption{Clinical-event components of the RECUR algorithm by cancer stage.}
\label{tab:clinical-events-by-stage}
\small
\resizebox{\textwidth}{!}{%
\begin{tabular}{lccccc}
\toprule
 & Overall & I & II & III & $p$-value \\
 & (N = 3{,}336) & (N = 2{,}200) & (N = 847) & (N = 289) & \\
\midrule
No. of secondary malignant neoplasm & & & & & $<0.001$ \\
\quad 0 & 2{,}821 (84.6\%) & 1{,}962 (89.2\%) & 669 (79.0\%) & 190 (65.7\%) & \\
\quad 1 & 36 (1.1\%) & 19 (0.9\%) & 9 (1.1\%) & 8 (2.8\%) & \\
\quad 2+ & 232 (7.0\%) & 77 (3.5\%) & 96 (11.3\%) & 59 (20.4\%) & \\
\quad Missing & 247 (7.4\%) & 142 (6.5\%) & 73 (8.6\%) & 32 (11.1\%) & \\
\addlinespace
No. of chemotherapy & & & & & 0.216 \\
\quad 0 & 258 (7.7\%) & 179 (8.1\%) & 53 (6.3\%) & 26 (9.0\%) & \\
\quad 1 & 235 (7.0\%) & 166 (7.5\%) & 53 (6.3\%) & 16 (5.5\%) & \\
\quad 2+ & 2{,}596 (77.8\%) & 1{,}713 (77.9\%) & 668 (78.9\%) & 215 (74.4\%) & \\
\quad Missing & 247 (7.4\%) & 142 (6.5\%) & 73 (8.6\%) & 32 (11.1\%) & \\
\addlinespace
No. of radiation therapy & & & & & $<0.001$ \\
\quad 0 & 1{,}212 (36.3\%) & 853 (38.8\%) & 278 (32.8\%) & 81 (28.0\%) & \\
\quad 1 & 680 (20.4\%) & 484 (22.0\%) & 146 (17.2\%) & 50 (17.3\%) & \\
\quad 2+ & 1{,}197 (35.9\%) & 721 (32.8\%) & 350 (41.3\%) & 126 (43.6\%) & \\
\quad Missing & 247 (7.4\%) & 142 (6.5\%) & 73 (8.6\%) & 32 (11.1\%) & \\
\addlinespace
No. of hospice & & & & & $<0.001$ \\
\quad 0 & 2{,}939 (88.1\%) & 1{,}999 (90.9\%) & 715 (84.4\%) & 225 (77.9\%) & \\
\quad 1 & 38 (1.1\%) & 19 (0.9\%) & 11 (1.3\%) & 8 (2.8\%) & \\
\quad 2+ & 112 (3.4\%) & 40 (1.8\%) & 48 (5.7\%) & 24 (8.3\%) & \\
\quad Missing & 247 (7.4\%) & 142 (6.5\%) & 73 (8.6\%) & 32 (11.1\%) & \\
\addlinespace
No. of mastectomy & & & & & $<0.001$ \\
\quad 0 & 2{,}960 (88.7\%) & 2{,}006 (91.2\%) & 724 (85.5\%) & 230 (79.6\%) & \\
\quad 1 & 60 (1.8\%) & 30 (1.4\%) & 22 (2.6\%) & 8 (2.8\%) & \\
\quad 2+ & 69 (2.1\%) & 22 (1.0\%) & 28 (3.3\%) & 19 (6.6\%) & \\
\quad Missing & 247 (7.4\%) & 142 (6.5\%) & 73 (8.6\%) & 32 (11.1\%) & \\
\addlinespace
No. of high cost imaging & & & & & $<0.001$ \\
\quad 0 & 1{,}062 (31.8\%) & 814 (37.0\%) & 187 (22.1\%) & 61 (21.1\%) & \\
\quad 1 & 370 (11.1\%) & 258 (11.7\%) & 89 (10.5\%) & 23 (8.0\%) & \\
\quad 2+ & 1{,}657 (49.7\%) & 986 (44.8\%) & 498 (58.8\%) & 173 (59.9\%) & \\
\quad Missing & 247 (7.4\%) & 142 (6.5\%) & 73 (8.6\%) & 32 (11.1\%) & \\
\addlinespace
No. of observation encounters & & & & & 0.139 \\
\quad 0 & 2{,}938 (88.1\%) & 1{,}971 (89.6\%) & 728 (86.0\%) & 239 (82.7\%) & \\
\quad 1 & 52 (1.6\%) & 32 (1.5\%) & 14 (1.7\%) & 6 (2.1\%) & \\
\quad 2+ & 99 (3.0\%) & 55 (2.5\%) & 32 (3.8\%) & 12 (4.2\%) & \\
\quad Missing & 247 (7.4\%) & 142 (6.5\%) & 73 (8.6\%) & 32 (11.1\%) & \\
\bottomrule
\end{tabular}%
}
\end{table}
\normalsize

\clearpage
\begin{figure}[!htbp]
    \centering
    \includegraphics[width=0.92\textwidth]{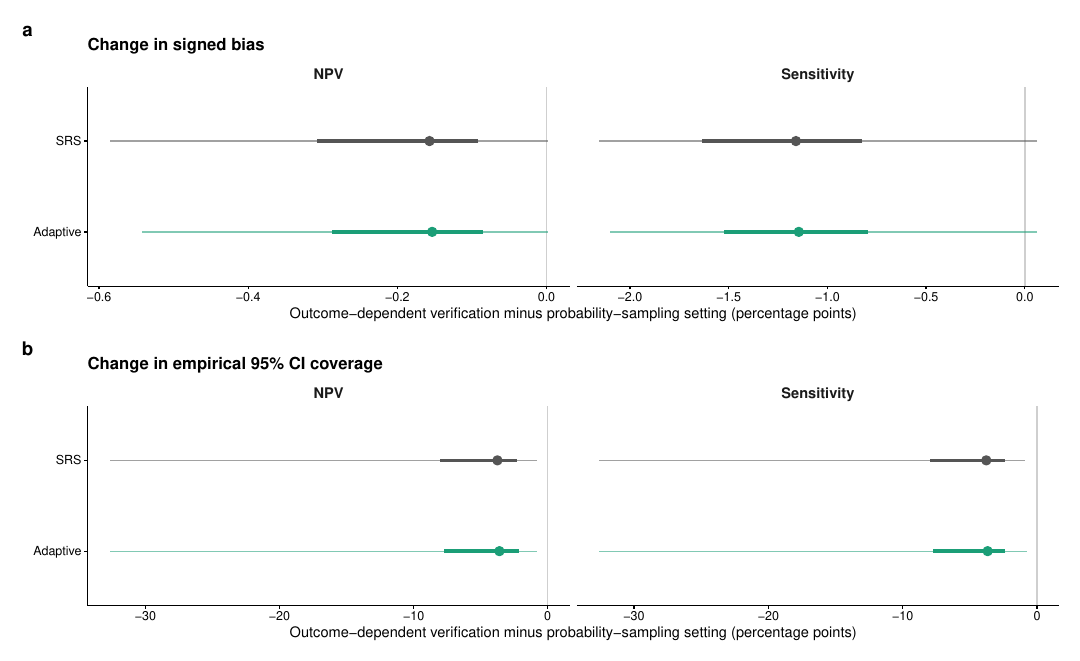}
    \caption{Sensitivity to outcome-dependent verification. Changes are calculated as the outcome-dependent verification scenario minus the primary probability-sampling scenario across 162 matched simulation settings. Points show medians, thick intervals show interquartile ranges, and thin intervals show 5th--95th percentile ranges.}
    \label{fig:outcome_dependent_verification}
\end{figure}

\end{document}